\documentclass[letter]{IEEEtran}
\usepackage{amsmath}
\usepackage{float}
\usepackage{graphicx}
\usepackage{multirow}
\usepackage{booktabs}
\usepackage{amsfonts,amssymb} 
\usepackage{algorithm}
\usepackage{algorithmic}

\ifCLASSINFOpdf
\else
\fi
\begin{document}
\title{Enhanced LMMSE Estimation Capable of Selecting Parameters}
% for OFDM Systems?

% searching for promotion, with Parameter Comparison


\author{\IEEEauthorblockN{Kai Mei\IEEEauthorrefmark{1},
Jun Liu\IEEEauthorrefmark{1},
Xiaoran Liu\IEEEauthorrefmark{1},
Jun Xiong\IEEEauthorrefmark{1}, 
Xiaoying Zhang\IEEEauthorrefmark{1}, and Jibo Wei\IEEEauthorrefmark{1}}\\
\IEEEauthorblockA{\IEEEauthorrefmark{1}School of Electronic Science and Engineering,
National University of Defense Technology, \\410073, Changsha, China}
% <-this % stops an unwanted space
\thanks{Manuscript received X X, 2019; revised X X, 2019.}}

%\markboth{IEEE Communication Transactions,~Vol.~X, No.~X, X~2019}%
%{Mei \MakeLowercase{\textit{et al.}}: An Enhanced LMMSE Channel Estimation Algorithm in OFDM System}




% for Transactions on Magnetics papers, we must declare the abstract and
% index terms PRIOR to the title within the \IEEEtitleabstractindextext
% IEEEtran command as these need to go into the title area created by
\IEEEtitleabstractindextext{ %
\begin{abstract}

In the linear minimum mean square error (LMMSE) estimation for orthogonal frequency division multiplexing (OFDM) systems, the problem about the determination of the algorithm's parameters, especially those related with channel frequency response (CFR) correlation, has not been readily solved yet. Although many approaches have been proposed to determine the statistic parameters, it is hard to choose the best one within those approaches in the design phase, since every approach has its own most suitable application conditions and the real channel condition is unpredictable. In this paper, we propose an enhance LMMSE estimation capable of selecting parameters by itself. To this end, sampled noise MSE is first proposed to evaluate the practical performance of interpolation. Based on this evaluation index, a novel parameter comparison scheme is proposed to determine the parameters which can endow LMMSE estimation best performance within a parameter set. After that, the structure of the enhanced LMMSE is illustrated, and it is applied in OFDM systems. Besides, the issues about theoretical analysis on accuracy of the parameter comparison scheme, the parameter set design and algorithm complexity are explained in detail. At last, our analyses and performance of the proposed estimation method are demonstrated by simulation experiments.

\end{abstract}

}

\maketitle

\IEEEdisplaynontitleabstractindextext

% \fi
\IEEEpeerreviewmaketitle

\section{Introduction}
\IEEEPARstart{O}{rthogonal} frequency division multiplexing (OFDM) is the enabling technology for numerous wireless communication standards, such as IEEE 802.11ac, IEEE 802.16, long term evolution (LTE), and 5G \cite{dobre2011second,6111235}. Channel estimation is one of the key issues in OFDM systems \cite{8743406,8904058,8903270}. The pilot assisted estimation has lower complexity and better performance and is widely used in practical systems. Among the pilot assisted estimation methods, minimum mean square error (MMSE) estimation has the best performance, but its computational complexity is much high \cite{van1995channel}. To reduce the complexity, the simplified MMSE estimation, called low-rank linear minimum mean square error (LMMSE) estimation, is proposed in \cite{501446}. In LMMSE estimation, statistic information about channel frequency response (CFR) correlation and signal-to-noise ratio (SNR) is required. When prior information about channel statistics cannot be provided, the problem about how to determine the statistic parameters in LMMSE estimation should be solved. There are many methods dealing with this problem, and it is hard to say which method is the best.

Among those methods, the simplest one is to predetermine the parameters in LMMSE estimation regardless of the statistics of real channel. With the predetermined parameters, LMMSE estimation will have robust performance in various channel conditions \cite{501446}. To aquire statistic information of wireless channel timely, many approaches have been proposed to estimate the SNR and CFR correlation. The SNR can be readily estimated, and the performance is close to the Cramer-Rao lower bound (CRLB) at high SNR \cite{SNRRen}. In contrast, the estimation of CFR correlation is difficult, because it is not a single parameter like SNR, but a function of frequency space. 

CFR correlation can be either directly estimated or indirectly obtained using the estimates of power delay profile (PDP) \cite{SavauxLMMSE}, since CFR correlation and PDP are a Fourier transform pair. CFR correlation can be directly estimated through averaging over the correlation between the estimates of CFR \cite{marques2005pilot}. The performance can be improved by iteratively estimating the noise variance and CFR correlation \cite{savaux2013application}. However, this kind of method requires a large number of pilot symbols. Otherwise, the estimation error of CFR correlation may cause significant performance loss for LMMSE estimation. When the coherence time is short or only a small number of pilots are transmitted, the methods that obtain CFR correlation based on PDP are better choices. One of such methods is proposed in \cite{zhou2009fast}. The authors utilize the sparsity of channel in time domain for the estimation of CFR correlation. In fact, this method is equivalent to estimating the PDP firstly and then applying FFT on PDP to obtain CFR correlation. Furthermore, the estimation of PDP can be simplified by using the approximate PDP. In \cite{YucekPDP}, PDP is determined by root-mean-squared (rms) delay and maximum excess delay. A common problem in those indirect CFR correlation estimation methods is that the employed PDP model may mismatch the actual channel. The modeling error will influence the overall channel estimation performance. The methods for the estimation of CFR correlation have their own suitable application conditions. But the real channel condition cannot be predicted, and thus it is hard to choose between those methods at the design phase.

In this paper, we try to enable the LMMSE estimator to select the statistic parameters by itself. In this way, CFR correlation obtained using different approaches can be compared and then the most suitable one for the real channel is chosen to perform LMMSE estimation. To this end, a novel scheme to choose the statistic parameters within a parameter set for LMMSE estimation is proposed. This scheme is based on the fact that LMMSE estimation and MMSE interpolation need the same channel statistic information, which can be tested using MMSE interpolation before performing LMMSE estimation. The performance of the parameter comparison scheme is analyzed, and fuzzy bound is proposed to describe the performance of the parameter comparison scheme. Fuzzy bound indicates the minimum performance difference that the scheme can distinguish at a certain confidence level. The confidence level is represented by false comparison possibility upper bound. Based on the parameter comparison scheme, an enhanced LMMSE estimator with parameter comparison ability is proposed. Then, the enhanced LMMSE estimation is applied in OFDM systems which employ the block pilot arrangement. To accomplish the research, the issues about parameter set design and computational complexity analysis are discussed. We design the set of parameter candidates under two types of conditions. One is that prior information is available, and the other is that no prior information can be provided. Besides, a simplified algorithm of much lower complexity is proposed, and the simulation result shows that the performance loss caused by the simplification is much small. Extensive computer simulation experiments are conducted to verify the theoretical analyses and performance of proposed estimator. The comparison with the conventional LMMSE estimation methods exhibits the superiority of the proposed estimator.

\emph{Notations}: We use boldface small letters and capital letters to denote vectors and matrices respectively. $\mathbb{E}\left[  \cdot  \right]$, $\mathbb{D}\left[  \cdot  \right]$, ${\mathcal{CN}\left( \cdot \right)}$, ${\left\|  \cdot  \right\|_2}$, and ${\left(  \cdot  \right)^{ - 1}}$ represent the expectation, the variance, the complex Gaussian distribution, the Euclidean norm, and the inversion, respectively. superscript T and H represent transpose and Hermitian transpose respectively.

\section{Enhanced LMMSE Estimation}
In this section, a novel statistic parameter comparison scheme is employed in LMMSE estimation to improve the quality of statistic parameters used in the algorithm. First, we introduce LMMSE estimation and MMSE interpolation. Then, we interpret the parameter comparison scheme and analyze its accuracy. Finally, an enhanced LMMSE estimation method is described.
 
\subsection{ LMMSE Estimation and MMSE Interpolation}
\label{subsec.1}

Consider that there is a sequence of stationary rough estimates ${{\bf{\hat h}}_{{\rm{LS}}}}$ to be filtered, which can be modelled as
\begin{equation}
{{\bf{\hat h}}_{{\rm{LS}}}} = {\bf{h}} + {\bf{n}},
\end{equation}
where ${\bf{h}}$ contains the true value of the sequence and ${\bf{n}}$ is a white complex Gaussian noise vector with variance $\sigma _{{\rm{LS}}}^2$. In channel estimation, ${{\bf{\hat h}}_{{\rm{LS}}}}$ is the LS estimation of channel response \cite{501446} and $\sigma _{{\rm{LS}}}^2$ is the mean square error (MSE) of LS estimation. ${\bf{h}}$ is assumed to be normalized, i.e. $\mathbb{E}\left[ hh^* \right] = 1$, where $h$ is an arbitrary element in ${\bf{h}}$. ${\bf{h}}$ is assumed to be subject to zero mean complex Gaussian distribution, i.e. $h \sim {\cal{CN}}\left( {0,1} \right)$.

To eliminate the influence of noise, the correlation between the elements of the sequence can be exploited in LMMSE estimation \cite{501446}:
\begin{equation}
\label{equ.mmse}
{{\bf{\hat h}}_{{\rm{LMMSE}}}} = {{\bf{R}}_{{\bf{hh}}}}{\left( {{{\bf{R}}_{{\bf{hh}}}} + \sigma _{{\rm{LS}}}^2{\bf{I}}} \right)^{ - 1}}{{\bf{\hat h}}_{{\rm{LS}}}},
\end{equation}
where ${{\bf{R}}_{{\bf{hh}}}}$ is the autocorrelation matrix of $\bf{h}$, and ${\bf{I}}$ is an identity matrix.

We denote the $i\rm{th}$ element of $\bf{h}$ as $h_i$. Through interpolating, the estimate of $h_i$ can be obtained using the rest elements in ${{\bf{\hat h}}_{{\rm{LS}}}}$, and we use ${{\bf{\hat h}}_{{i\rm{Ex}}\_ {\rm{LS}}}}$ to represent the vector containing those elements. The MMSE interpolation of $h_i$ using ${{\bf{\hat h}}_{{i\rm{Ex}}\_ {\rm{LS}}}}$ is given by \cite{DongInter}
\begin{equation}
\label{equ.int1}
{\hat h_{i\_{\rm{Int}}}} = {{\bf{r}}_{{h_i}{{\bf{h}}_{i{\rm{Ex}}}}}}{\left( {{{\bf{R}}_{{{\bf{h}}_{i{\rm{Ex}}}}{{\bf{h}}_{i{\rm{Ex}}}}}} + \sigma _{{\rm{LS}}}^2{\bf{I}}} \right)^{ - 1}}{{\bf{\hat h}}_{i{\rm{Ex}}\_{\rm{LS}}}},
\end{equation}
where ${{\bf{h}}_{i\rm{Ex}}}$ contains the true value of  ${{\bf{\hat h}}_{{i\rm{Ex}}\_ {\rm{LS}}}}$. ${{\bf{r}}_{{h_i}{{\bf{h}}_{i\rm{Ex}}}}}$ is the correlation matrix between ${h_i}$ and ${{\bf{h}}_{i\rm{Ex}}}$ and ${{\bf{R}}_{{{\bf{h}}_{i\rm{Ex}}}{{\bf{h}}_{i\rm{Ex}}}}}$ is the autocorrelation matrix of ${{\bf{h}}_{i\rm{Ex}}}$.

Both LMMSE estimation and MMSE interpolation require noise variance and correlation information. We use ${\pmb{\theta }} = \left\{ {r\left( 0 \right),r\left( 1 \right),...,r\left( {K - 1} \right),{\sigma _{{\rm{LS}}}^2}} \right\}$ to represent the vector that contains those statistic parameters, where $r\left( {\Delta k} \right)$ is the correlation of ${\bf{h}}$ spaced by ${\Delta k}$, i.e. $r\left( {\Delta k} \right) = E\left[ {{h_{k + \Delta k}}{{\left( {{h_k}} \right)}^ * }} \right]$. With ${\pmb{\theta }} $, the autocorrelation matrix ${{\bf{R}}_{{\bf{hh}}}}$ required in LMMSE estimation can be generated as 
\begin{equation}
\label{equ.Rconstruct}
\begin{aligned}
{{\bf{R}}_{{\bf{hh}}}}\left[ {i,j} \right] = \left\{ \begin{matrix}
  r\left( {i - j} \right),\;\;\;\; i \ge j \hfill \cr 
  {\left( {r\left( {j - i} \right)} \right)^ * }.\; i < j \hfill \cr \end{matrix}  \right.
\end{aligned}
\end{equation}

\begin{figure}[!htb]
\begin{centering}
\includegraphics[scale=0.9]{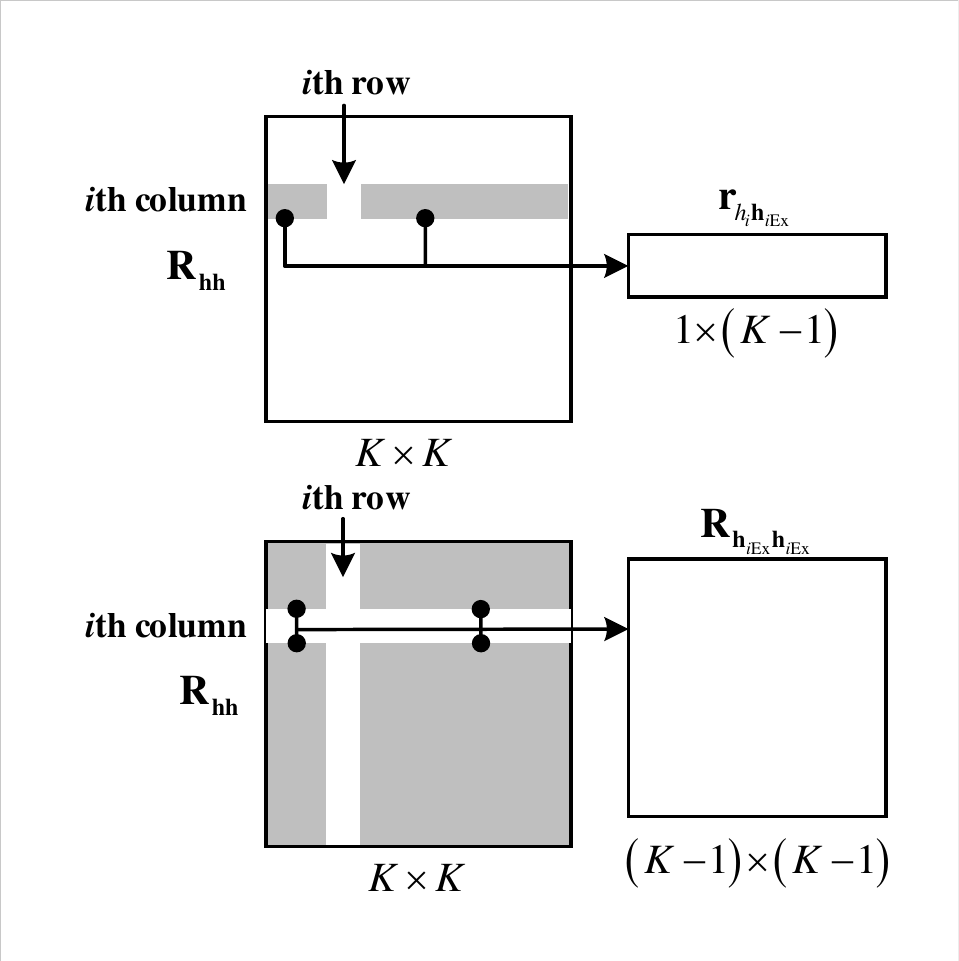}
\par\end{centering}
\caption{Sketch diagram showing that correlation matrices in MMSE interpolation are part of autocorrelation matrix.}
\label{fig.CorMat}
\end{figure}

${{\bf{r}}_{{h_i}{{\bf{h}}_{i\rm{Ex}}}}}$ and ${{\bf{R}}_{{{\bf{h}}_{i\rm{Ex}}}{{\bf{h}}_{i\rm{Ex}}}}}$ are part of ${{\bf{R}}_{{\bf{hh}}}}$ as shown in Fig. \ref{fig.CorMat}. Therefore, with ${\pmb{\theta }}$,  LMMSE estimation and MMSE interpolation can be performed on ${{\bf{\hat h}}_{{\rm{LS}}}}$. We use ${{\bf{\hat h}}_{{\rm{LMMSE}}\left| {\pmb{\theta }} \right.}}$ and ${\hat h_{i\_{\rm{Int}}\left| {\pmb{\theta }} \right.}}$ to represent the LMMSE estimation and MMSE interpolation using the parameters in ${\pmb{\theta }}$ respectively. 

\subsection{ Parameter comparison scheme}

Suppose that there is a set of parameter vectors $\Omega = \left\{ {{\pmb{\theta} _1},...,{\pmb{\theta} _N}} \right\}$ to be compared. It is hard to directly determine which parameter vector is closest to statistics of ${{\bf{\hat h}}_{{\rm{LS}}}}$, so we evaluate the parameter vectors by the performance of MMSE interpolations in which the corresponding parameter vector is used. MSE is often used to evaluate the performance of interpolation. However, in practical systems, MSE cannot be obtained, so we come up with a substitution for MSE, which we call as sampled noise MSE. The sampled noise MSE for ${\pmb{\theta} _n}$ is expressed as 
\begin{equation}
\label{equ.metric}
\xi_n  = {1 \over K}\sum\limits_k {{{\left| {{{\hat h}_{k\_{\rm{Int}}\left| {\pmb{\theta }_n} \right.}} - {{\hat h}_{k\_{\rm{LS}}}}} \right|}^2}},
\end{equation}   
where $K$ is the dimension of ${\bf{h}}$. ${{\hat h}_{k\_{\rm{Int}}\left| {\pmb{\theta }_n} \right.}}$  is MMSE interpolation based on the parameters in ${\pmb{\theta} _n}$ and specifically expressed as (\ref{equ.int1}). The differences between the sampled noise MSE and MSE lie in two aspects: first, MSE is the expectation of estimation error and sampled noise MSE is the sample  average of estimation error; second, the true values are used to calculate estimation error in MSE, while it is the rough estimates in ${{\bf{\hat h}}_{{\rm{LS}}}}$ that are used to calculate estimation error in sampled noise MSE.

$\mathbb{E}\left[  {{\left| {{{\hat h}_{k\_{\rm{Int}}}} - {{\hat h}_{k\_{\rm{LS}}}}} \right|}^2} \right]$ is derived in Appendix \ref{IndexVariance} and it equals $\sigma _{{\text{MSE}}}^2 + \sigma _{\rm{LS}}^2$, where $\sigma _{{\text{MSE}}}^2$ is the MSE of interpolation. When $K$ is big enough, $\xi$ will approach $\sigma _{{\text{MSE}}}^2+\sigma _{\rm{LS}}^2$. We call $\sigma _{{\text{MSE}}}^2+\sigma _{\rm{LS}}^2$ as noise MSE, and thus we call $\xi$ as sampled noise MSE. For different interpolation methods, the value of noise MSE is mainly determined by its MSE value, since $\sigma _{\rm{LS}}^2$ is the value of LS estimation and not influenced by the interpolation methods. Thus, the performance index $\xi$ of a good parameter vector tends to be lower than that of a bad one. 

To select the best parameter vector from $\Omega$, we first use the vectors to perform MMSE interpolation respectively. Then, we evaluate the performance of the interpolations using the index $\xi_n$ and the parameter vector ${\pmb{\theta} _*}$ corresponding to the best interpolation is seen as the most suitable one for filtering ${{\bf{\hat h}}_{{\rm{LS}}}}$. The parameter comparison algorithm is illustrated in Algorithm \ref{tab:PCA}.
\begin{algorithm}[htbp]
\caption{parameter comparison algorithm}
\label{tab:PCA}
\centering
\begin{algorithmic}[1]	
\REQUIRE ~~\\
The set of parameter vectors $\Omega $, noisy sequence ${{\bf{\hat h}}_{{\rm{LS}}}}$;
\ENSURE ~~\\
The best parameter vector, ${\pmb{\theta} _*}$;
    \FOR{$n = 1,2,...,N$,}
    \STATE $\rm{sum}=0$;
    \FOR{$k = 1,2,...,K$,}
    \STATE generate ${{\bf{R}}_{{h_k}{{\bf{h}}_{k\rm{Ex}}}}}$, ${{\bf{R}}_{{{\bf{h}}_{k\rm{Ex}}}{{\bf{h}}_{k\rm{Ex}}}}}$, SNR based on ${\pmb{\theta} _n}$;
    \STATE calculate ${\rm{err}} = {\left| {{{\hat h}_{k\_{\rm{Int}}}} - {{\hat h}_{k\_{\rm{LS}}}}} \right|^2}$ according to (\ref{equ.int1});
    \STATE ${\rm{sum = sum + err}}$; 
    \ENDFOR
    \STATE ${\xi _n} = {\rm{sum/}}K$;
    \ENDFOR
    \STATE ${\pmb{\theta} _*}  = {\pmb{\theta} _i},\;i = \mathop {\arg }\limits_n \min {\xi _n}$.
\end{algorithmic}
\end{algorithm}

\subsection{Fuzzy bound of parameter comparison}
\label{PerfAna}

Since $\xi$ is the sampled noise MSE, the value of $\xi$ is random and the evaluation index $\xi$ of a good parameter vector may be higher than that of a worse one. In that case, wrong comparison result will be drawn. It is intuitive that wrong comparison occurs at higher possibility when the MSE of the interpolations using two parameter vectors are close. Thus, there exists an MSE difference bound, and when the MSE difference of the two interpolations is lower than the bound, the parameter comparison scheme will be invalid. We call that bound as fuzzy bound of parameter comparison. In the following part of this Subsection, we will analyze the fuzzy bound theoretically.

Suppose that there are two parameter vectors ${\pmb{\theta} _1}$ and ${\pmb{\theta} _2}$ to be compared for a certain noisy sequence ${{\bf{\hat h}}_{{\rm{LS}}}}$. We assume that for this sequence, ${\pmb{\theta} _1}$ is better than ${\pmb{\theta} _2}$ and MSE of the interpolation based on ${\pmb{\theta} _1}$ and ${\pmb{\theta} _2}$ are $\sigma _1^2$ and $\sigma _2^2$  respectively, where $\sigma _{\rm{MSE}1}^2$ is lower than $\sigma _{\rm{MSE}2}^2$ and $\sigma _{\rm{MSE}2}^2 = \sigma _{\rm{MSE}1}^2 + {\Delta _{{\rm{MSE}}}}$. Denote the noise MSE of the interpolation based on ${\pmb{\theta} _1}$ and ${\pmb{\theta} _2}$ as $\sigma _1^2$ and $\sigma _2^2$respectively, i.e. $\sigma _1^2=\sigma _{\rm{MSE}1}^2+\sigma _{\rm{LS}}^2$ and $\sigma _2^2=\sigma _{\rm{MSE}2}^2+\sigma _{\rm{LS}}^2$. Based on two	 assumptions, the false comparison possibility $\varepsilon $ is derived in Appendix \ref{Possibility}, and it is influenced by two factors: length of sequence $K$ and scaled performance difference ${\Delta _{{\rm{MSE}}}}/\sigma _2^2$. Let $\alpha = {\Delta _{{\rm{MSE}}}}/\sigma _2^2$ and the expression of $\varepsilon $ can be simplified as 	
\begin{equation}
\label{equ.EasyPossib}
\begin{aligned}
   \varepsilon  &= {f_{{\rm{FP}}}}\left( {\alpha ,K} \right)  \cr 
  &  = \mathbb{E}\left[ {F\left( {\left( {1 - \alpha } \right){{\cal{X}} _{2K}}} \right)} \right],
\end{aligned}
\end{equation}   
where ${f_{{\rm{FP}}}}\left(  \cdot  \right)$ represents some function of two variables and $F\left(  \cdot  \right)$ is the cumulative density function (CDF) of a random variable. ${\cal{X}} _{2K}$ denotes a random variable which is subject to the chi-square distribution of $2K$ dimension.

\begin{figure}[!htb]
\begin{centering}
\includegraphics[scale=0.5]{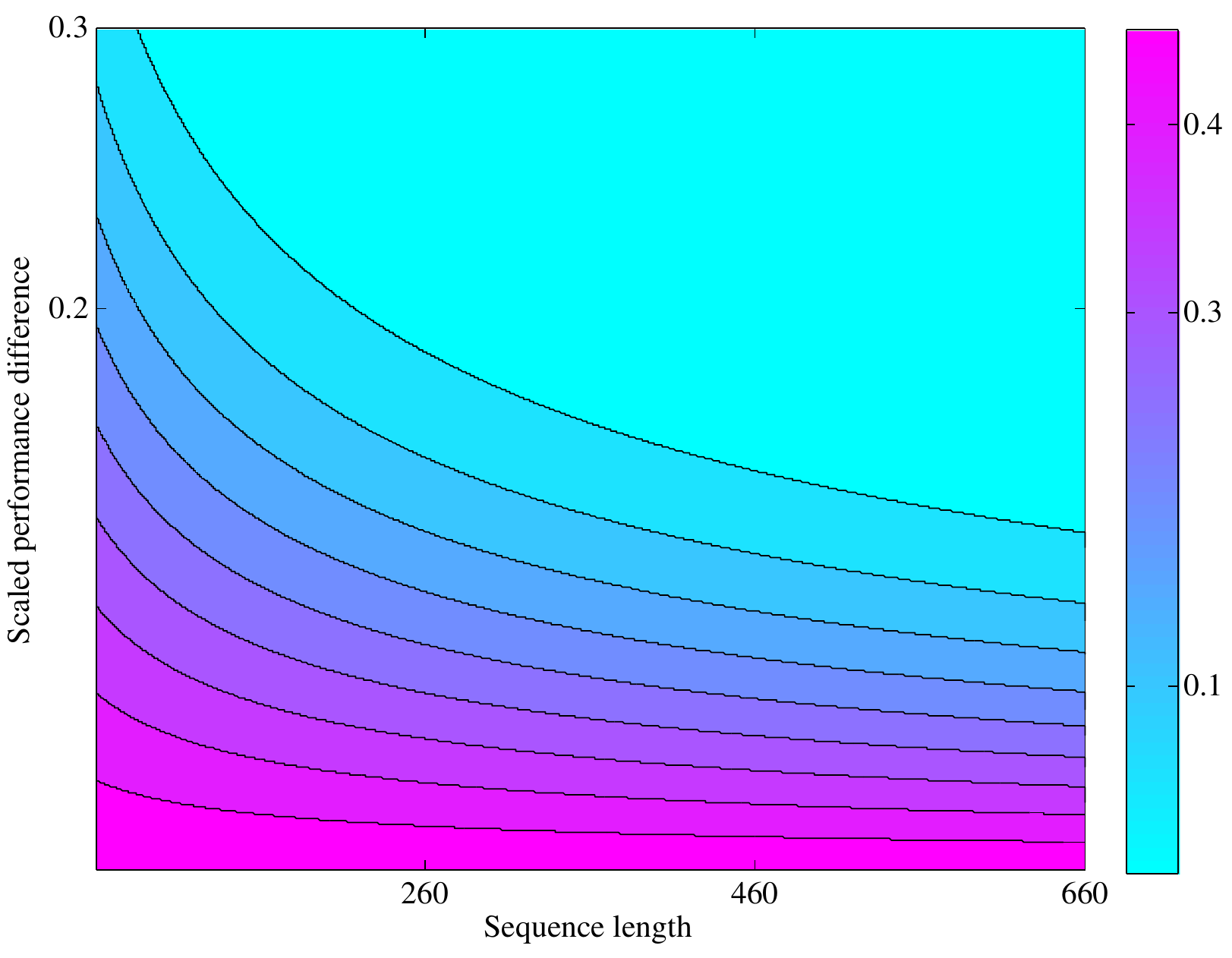}
\par\end{centering}
\caption{Error possibility varied with the sequence length $K$ and the scaled performance difference $\alpha$.}
\label{fig.3Dpossib}
\end{figure}

We plot $\varepsilon $ varied with $\alpha$ and $K$. From Fig. \ref{fig.3Dpossib}, we can see that the false comparison possibility $\varepsilon $ decreases with the increase of $\alpha$ and $\kappa $. It confirms the intuitive fact that the accuracy of parameter comparison will get higher with the increase of sequence length and parameters' performance difference. From this figure, two basic properties of the parameter comparison scheme can be deduced: when the sequence length is too small, the parameter comparison scheme cannot be applied; there are mainly three factors affecting the accuracy of parameter comparison: the length of sequence, performance difference and noise variance.

We define fuzzy bound as the scaled performance difference $\alpha$ when false comparison possibility equals ${{\varepsilon _0}}$, and denote it as $\mathcal{B}_{{\varepsilon _0}}$. When the sequence length is given, fuzzy bound $\mathcal{B}_{{\varepsilon _0}}$ can be predicted using (\ref{equ.EasyPossib}). Fuzzy bound describes the minimum performance difference with a false comparison possibility lower than ${{\varepsilon _0}}$ guaranteed. It reflects the distinguishing ability of parameter comparison scheme and can be regarded as the resolution of parameter comparison. Higher resolution stands for better comparison scheme.

The parameter comparison scheme will always bring performance promotion for LMMSE estimation on the average, although wrong comparison may frequently occur. This claim is  demonstrates as following. We assume that performance differences of the parameter vectors are above the fuzzy bound. Then, the inequality holds that $\Delta \ge \mathcal{B}_{{\varepsilon _0}}\left({2*{\varepsilon _0}-1}\right)$, where $\Delta$ represents the average of scaled performance change. ${{\varepsilon _0}}< {f_{{\rm{FP}}}}\left( {0,1} \right) = 0.5$. Thus, ${{\varepsilon _0}}<0.5$ always holds, and thus $\Delta$ is always negative. The decrease of MSE means the increase of performance. Therefore, the parameter comparison scheme will always bring performance promotion on the average sense. The performance promotion is finally determined by the quality of parameter vectors in the parameter set. The better performance that the parameter vectors may have, the more performance promotion the proposed method can provide.

\begin{figure}[!htb]
\begin{centering}
\includegraphics[scale=0.95]{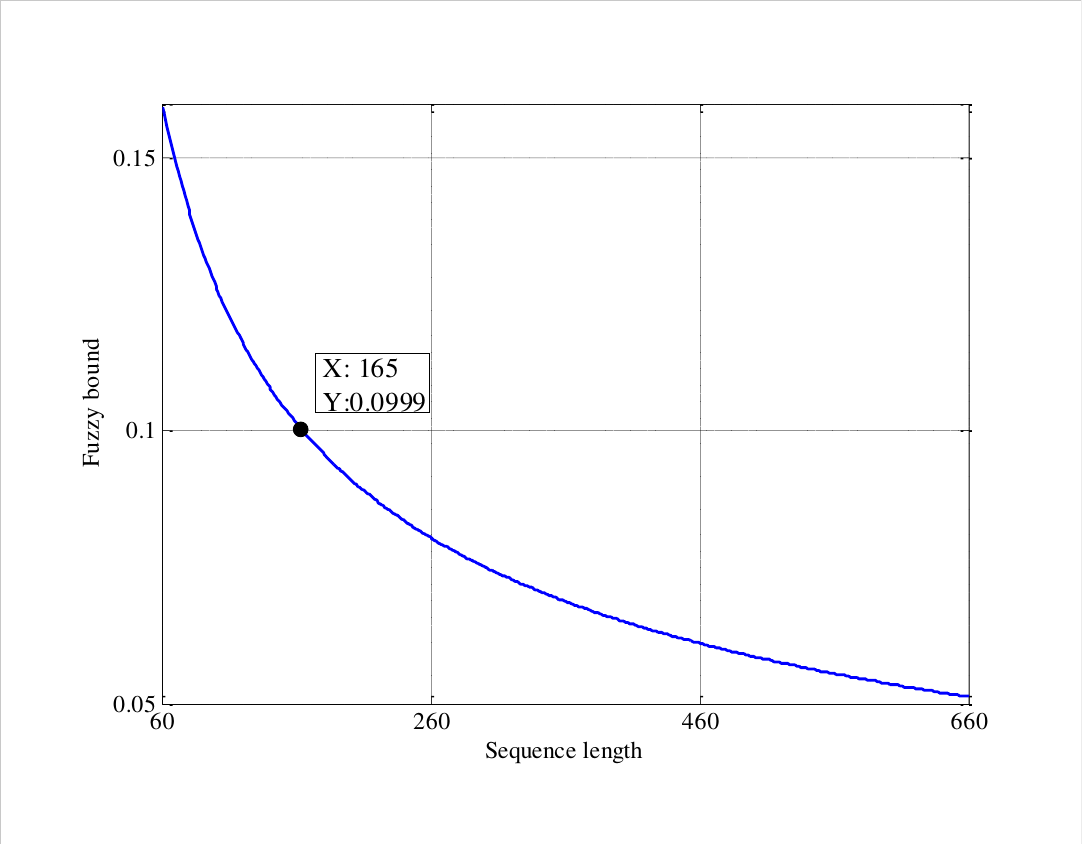}
\par\end{centering}
\caption{Fuzzy bound varied with the sequence length.}
\label{fig.FuzzyB}
\end{figure}

To display an example of fuzzy bound curve, we simply choose  ${{\varepsilon _0}}=0.25$. Then, we plot $\mathcal{B}_{0.25}$ varied with $K$. From Fig. \ref{fig.FuzzyB}, we can see that $\alpha$ is below 0.1 when the sequence length comes to 165. It shows that when the sequence length is above 165, the parameter comparison scheme has relatively high resolution.
  
\subsection{LMMSE Estimator with parameter comparison}

With prior information about the statistics of the sequence ${{\bf{\hat h}}_{{\rm{LS}}}}$, parameters in LMMSE algorithm can be determined directly. But prior information may mismatch the real situations, and to avoid the mismatch problem, the parameters can be obtained timely by estimation. However, when the data is limited, the estimation error of parameters may cause significant performance loss. In that case, robust parameters are good choice. These methods have their own suitable application conditions. However, the application conditions are normally hard to predict, and thus it is difficult to determine which method is better during the design of practical systems. In stead of choosing a method in advance, the parameter vector set $\Omega $ that contains the statistic parameters derived through different approaches, can be prepared for selection. 

\begin{figure}
\begin{centering}
\includegraphics[scale=0.7]{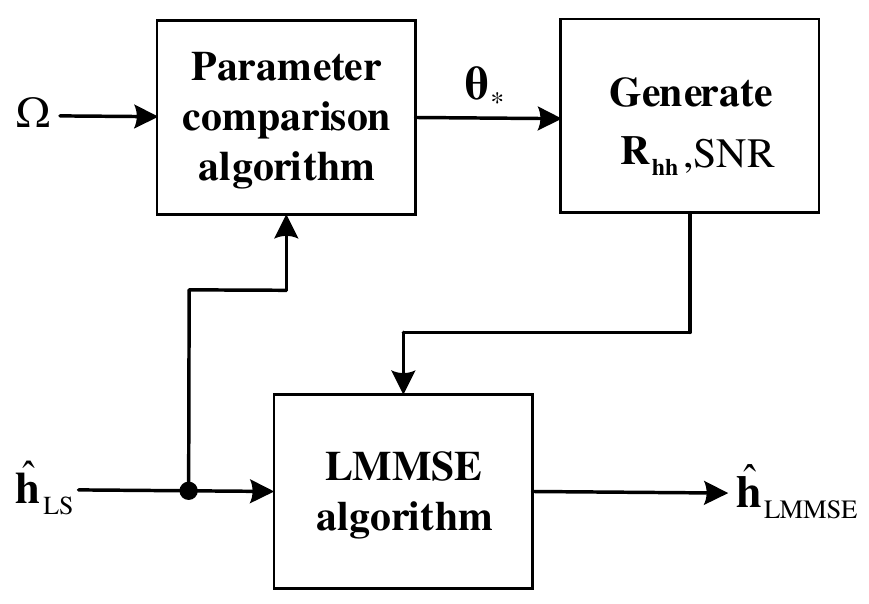}

\end{centering}
\caption{The sketch diagram of proposed LMMSE estimator with parameter comparison.}
\label{figest}
\end{figure}

After the rough estimates is obtained, the estimator first compares the candidates in $\Omega$ with the proposed parameter comparison algorithm. After the most suitable parameter ${\pmb{\theta} _*}$ vector for the sequence ${{\bf{\hat h}}_{{\rm{LS}}}}$ is selected,  the autocorrelation matrix ${{\bf{R}}_{{\bf{hh}}}}$ are then generated based on ${\pmb{\theta} _*}$. With ${{\bf{R}}_{{\bf{hh}}}}$ and SNR value, LMMSE algorithm can be performed to filter the noisy sequence. The estimator structure is shown in Fig. \ref{figest}.

From \ref{figest}, we can see that using the noise sequence itself can test the quality of statistic parameters in the filtering algorithm, since the correlation and noise variance information is already contained in the noisy sequence. When the data amount of the sequence is above 165, as explained in Section \ref{PerfAna}, the implicit statistic property of the noisy sequence can be exploited to examine the statistic parameters using our parameter comparison algorithm. With parameter comparison ability, the estimator can select the best statistic parameters derived through different approaches. The performance of LMMSE estimation therefore can be promoted through the additional parameter comparison operation.

\section{Enhanced LMMSE estimator for OFDM systems}

Many approaches have been proposed to determine the statistic parameters required in LMMSE estimation for OFDM systems. Instead of choosing one of those methods, we use those methods to generate different correlation vectors and then, through the parameter comparison algorithm, the enhanced LMMSE estimator selects the best parameter vector. Since SNR can be readily obtained in most OFDM systems, we simply assume it is known in this paper. So the enhanced LMMSE estimator for OFDM systems is designed to select correlation parameters. In this paper, the OFMD systems with block pilot arrangement is investigated, and thus correlation in frequency domain is exploited in LMMSE estimation. In this section, we first introduce the system model. Following that, the enhanced LMMSE estimator for OFDM system is illustrated. At last, the design of parameter set is explained in detail.

\subsection{System Model}

We consider OFDM systems under wide-sense stationary uncorrelated scattering (WSSUS) channel condition. The channel impulse response (CIR) is assumed to be constant during one OFDM symbol. The length of the cyclic prefix (CP) is assumed to be larger than the channel length. Therefore, the inter-carrier interference (ICI) caused by Doppler shift and inter-symbol interference caused by large delay are not considered in this paper. When the time and frequency synchronization are accurate, the received signal after CP remove and DFT operation can be written as 
\begin{equation}
\label{equ:chap2.28}
{\bf{y}} = {{\bf{X}}}{{\bf{h}}}^{\rm{f}} + {\bf{z}},
\end{equation}
where ${\bf{X}} = {\rm{diag}} \{ x_1,...,x_K\}$ is a diagonal matrix containing the transmitted signal. ${\bf{h}^{\rm{f}}}={[h_1^{\rm{f}},...,h_K^{\rm{f}}]}^T$ denotes the  channel frequency fading vector, and it is assumed to obey the complex Gaussian distribution, i.e. ${{\bf{h}^{\rm{f}}}{\sim}{\mathcal{CN}}\left( {{\bf{0}},{{\bf{R}}_{{\bf{h^{\rm{f}}h^{\rm{f}}}}}}} \right)}$. ${\bf{z}}={[z_1,...,z_K]}^T$ is the white Gaussian noise vector, i.e. ${{\bf{z}}{\sim}{\mathcal{CN}}\left( {{\bf{0}},\sigma _{\rm{n}}^2{\bf{I}}} \right)}$, where ${\bf{I}}$ denotes the identity matrix and $\sigma _{\rm{n}}^2$ is variance of noise. $\bf{z}$ is independent of ${\bf{h}}^{\rm{f}}$. 

In practical systems, symbol timing offset (STO) and carrier frequency offset (CFO) may exist. When STO is present, the received signal would be \cite{AthaudageEnhance}
\begin{equation}
\label{equ.STOmodel}
y_k = {x_k}h_k^{\text{f}}{e^{ - \frac{{j2\pi k\theta }}{K}}} + {z_k},
\end{equation}
where $\theta$ is the timing error. The value of $\theta$ is normally negative and the timing point is within CP. Let $\tilde h_k^{\text{f}} = h_k^{\text{f}}{e^{ - \frac{{j2\pi k\theta }}{K}}}$ be the effective CFR. The correlation of effective CFR will be 
\begin{equation}
\label{STOcor}
\begin{aligned}
   {{\tilde r}^{\rm{f}}}\left( {\Delta k} \right) &= {\mathbb{E}} \left[ {\tilde h_{k+\Delta k}^{\rm{f}}{{\left( {\tilde h_k^{\rm{f}}} \right)}^ * }} \right]  \cr 
  & = {\mathbb{E}} \left[ {h_{k{\rm{ + }}\Delta k}^{\rm{f}}{e^{ - {{j2\pi \left( {k+\Delta k} \right)\theta } \over K}}}{{\left( {h_k^{\rm{f}}{e^{ - {{j2\pi k\theta } \over K}}}} \right)}^ * }} \right]  \cr 
  & ={e^{ - {{j2\pi \theta \left( {k+\Delta k-k} \right)} \over K}}} {\mathbb{E}} \left[ {h_{k{\rm{ + }}\Delta k}^{\rm{f}}{{\left( {h_{k}^{\rm{f}}} \right)}^ * }} \right]  \cr 
  & ={e^{ - {{j2\pi \theta \Delta k} \over K}}}{r^{\rm{f}}}\left( {\Delta k} \right) ,
\end{aligned}
\end{equation}
where ${r^{\rm{f}}}\left( {\Delta k} \right)$ represents the correlation of real CFR. We can see that STO causes accumulated phase shifts among the subcarriers. The effective correlation will change as well. Therefore, the statistic property of  effective CFR will be different from the real CFR.

When CFO is present, the received signal can be modelled as
\begin{equation}
\label{equ.CFOmodel}
{y_k} = {x_k}h_k^{\rm{f}}{e^{{\phi}}} + {\tilde z_k},
\end{equation}
where $\phi$ denotes the common phase rotation on the CFR of the $m \rm{th}$ OFDM symbol. ${\tilde n_k} = n_k^{{\rm{ICI}}}+{n_k}$ represents the effective noise, where $n_k^{{\rm{ICI}}}$ stands for the ICI caused CFO and is normally assumed to be white Gaussian noise. Therefore, ${\tilde n_k}$ still obeys complex Gaussian distribution and has higher variance compared with ${n_k}$. The correlation of effective CFR is 
\begin{equation}
\label{CFOcor}
\begin{aligned}
   {{\tilde r}^{\rm{f}}}\left( {\Delta k} \right) &= {\mathbb{E}}  \left[ {\tilde h_{k + \Delta k}^{\rm{f}}{{\left( {\tilde h_k^{\rm{f}}} \right)}^*}} \right] \cr 
  & = {\mathbb{E}} \left[ {h_{k{\rm{ + }}\Delta k}^{\rm{f}}{e^{ - {\phi }}}{{\left( {h_k^{\rm{f}}{e^{ - {\phi }}}} \right)}^*}} \right]  \cr 
  & ={e^{ - {\phi }}}{e^{{\phi }}}{\mathbb{E}}\left[ {h_{k + \Delta k}^{\rm{f}}{{\left( {h_k^{\rm{f}}} \right)}^*}} \right]  \cr 
  & ={r^{\rm{f}}}\left( {\Delta k} \right).
\end{aligned}
\end{equation}

The effective correlation equals the real correlation. Thus, unlike STO, CFO will not influence the statistic property of CFR.

\subsection{Enhanced LMMSE estimator for OFDM}

In OFDM systems with the block pilot arrangement, the whole tones in a pilot symbol are used to estimate the CFR. The LS estimation of CFR is simple and can be expressed as
\begin{equation}
\hat {\bf h}_{\rm{LS}}^{\rm{f}}= {\bf X}^{-1}{\bf y}.
\end{equation}

We assume the pilot symbols are normalized. According to (\ref{equ.mmse}), the LMMSE estimation for the considered OFDM systems is
\begin{equation}
 {{\bf{\hat h}^{\rm{f}}}_{{\rm{LMMSE}}}} = {{\bf{R}}_{{\bf{h^{\rm{f}}h^{\rm{f}}}}}}{\left( {{{\bf{R}}_{{\bf{h^{\rm{f}}h^{\rm{f}}}}}} + \sigma _{\rm n}^2{\bf{I}}} \right)^{ - 1}}{{\bf{\hat h}}_{{\rm{LS}}}^{\rm{f}}}.
 \end{equation}
The LMMSE estimation in such OFDM system can be regarded as a filter which smooths out a noisy sequence of length $K$. In most OFDM systems, the number of subcarriers per symbol is normally 200 or more, so our proposed enhanced LMMSE estimation can be applied. 

\begin{figure}
\begin{centering}
\includegraphics[scale=0.6]{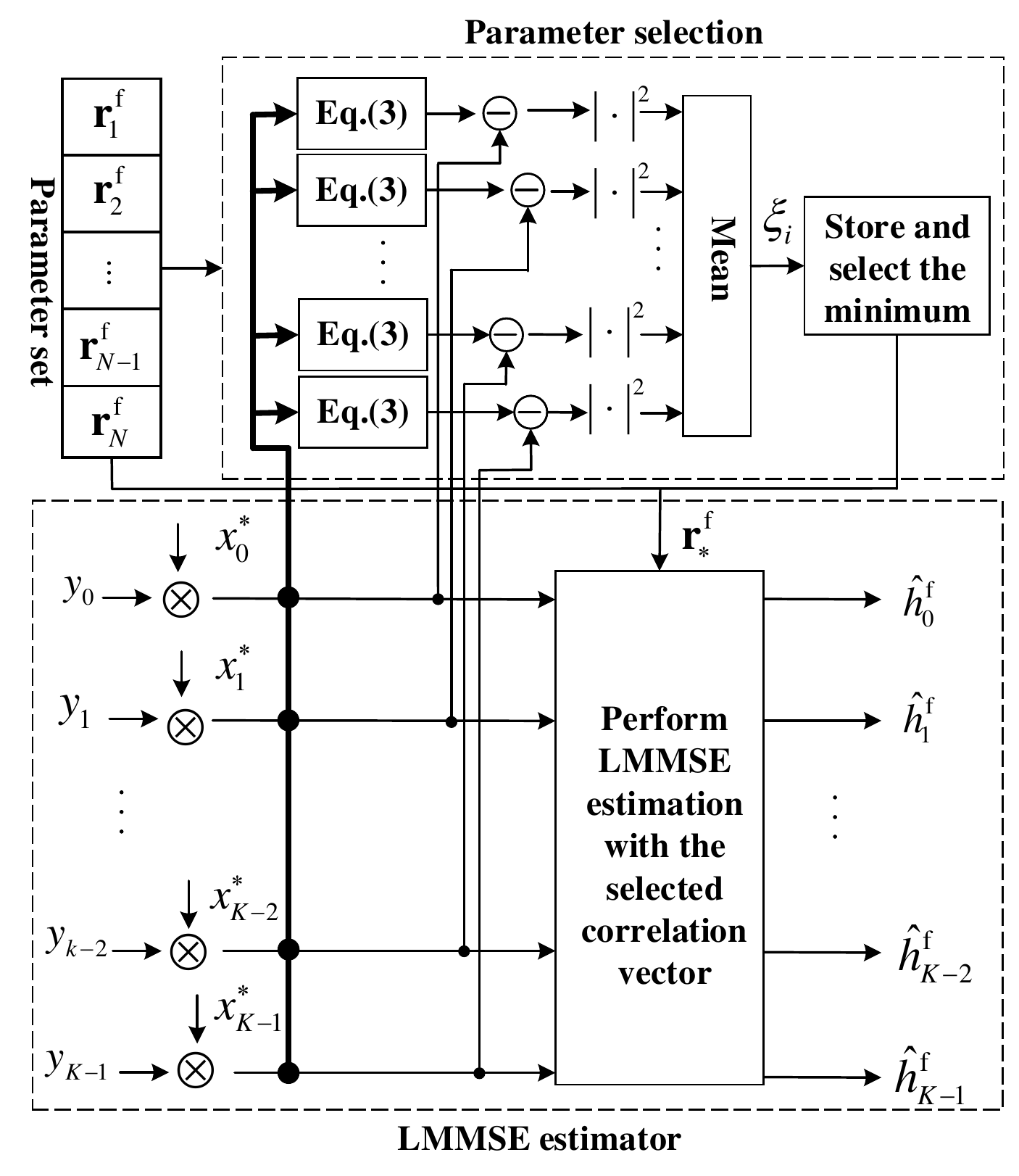}

\end{centering}
\caption{The sketch diagram of proposed LMMSE estimator with parameter comparison.}
\label{Fig.Est4OFDM}
\end{figure}

Referring to the estimator structure in Fig. \ref{figest}, the estimator for OFDM systems can be directly designed. We only consider the selection of correlation, so the noise variance should be excluded from the parameter vector ${\pmb{\theta }}$. To distinguish from the parameter vector in Section \ref{subsec.1}, we change the denotation of the parameter vector and denote ${\bf{r}}^{\rm{f}}$ as the parameter vector containing correlation only, i.e. ${\bf{r}}^{\rm{f}}=\left\{ {r^{\rm{f}} \left( 0 \right),r^{\rm{f}} \left( 1 \right),...,r^{\rm{f}} \left( {K - 1} \right)} \right\}$. It is also called as the CFR correlation vector. Therefore, the parameter set would be $\Omega  = \left\{ {{\bf{r}}_1^{\rm{f}},{\bf{r}}_2^{\rm{f}},...,{\bf{r}}_N^{\rm{f}}} \right\}$. The structure of enhanced LMMSE estimator for OFDM is shown in Fig. \ref{Fig.Est4OFDM}.

After enabling the estimator with parameter selection, we consider the design of parameter set. A good parameter set is crucial for the promotion of performance in our proposed estimator.  There are many approaches to determine the correlation of CFR, and the obtained correlation varies a lot with different approaches. Some methods predetermine the correlation which has robust performance under common channel conditions, some methods estimate the correlation timely and some methods obtain the CFR correlation through the estimation of PDP.  The derived CFR correlation through these methods can be divided into two types. One type is the robust correlation \cite{501446}. With the robust correlation, LMMSE estimation can be applied under wide range of channel conditions regardless of the real correlation. But the performance improvement over LS estimation is limited. The other type is the approximation of real correlation. With this kind of correlation, LMMSE estimation can achieve good performance when the employed correlation is close to real value, but the performance may decrease sharply when there are difference between the real correlation and the used one. These methods have their own suitable channel conditions and it is hard to say which one is the best before the system is put into use. A good design of parameter set is to contain the two types correlation. In this way, the enhanced estimator can explore for performance promotion as much as possible, and meanwhile an acceptable performance is guaranteed. In the next subsection, we will illustrate the design of parameter set in detail.

\subsection{Parameter set design}
\label{Sect.ParaSetDes}

We consider the design of parameter set under mainly two kinds of application scenarios: one is that information about CFR correlation can be provided through measurement, and the other is that no prior information is known. When the systems will used in changeless environments, such as in apartments, the CFR correlation is relatively stable. After channel measurement,  the correlation vectors that achieve good performance frequently can be derived through the analysis of the measurement data. Those correlation vectors can be used to construct the parameter set $\Omega$. To avoid huge performance degradation under unexpected channel conditions, robust correlation should be added into the parameter set as well. A most used robust correlation is the fast Fourier transform (FFT) of a uniform PDP. The PDP is described in (\ref{equ.PDP}).
\begin{equation}
\label{equ.PDP}
 \Gamma \left( \tau  \right) = \left\{ \begin{gathered}
   \frac{1}{\tau _{\rm{max}} },\; \tau \leq \tau _{\rm{max}} \hfill \\
  0, \; \; \; \; \; \;\;  \tau > \tau _{\rm{max}} \hfill \\ 
\end{gathered}  \right.
 \end{equation}
where $\tau _{\rm{max}}$ represents the possible maximum delay. It is often set as the length of CP. 

Note that STO influences the effective correlation. In the LMMSE estimation algorithm, the effective correlation ${{\tilde r}^{\rm{f}}}\left( {\Delta k} \right)$ should be used rather than ${r^{\rm{f}}}\left( {\Delta k} \right)$ \cite{AthaudageEnhance}. Therefore, STO should be considered in the design of parameter set. Denote ${\mathbf{\tilde r}}^{\text{f}}_{\theta}$ as the effective correlation vector under ${\theta}$ timing offset. ${\mathbf{\tilde r}}^{\text{f}}_{\theta}$ for all possible STO value can be added into $\Omega$. However, $\Omega$ will be too large, if ${\mathbf{\tilde r}}^{\text{f}}_{\theta}$ for all the possible STO are included. An alternative method is to incorporate the statistics of STO into the CFR correlation \cite{AthaudageEnhance}: 
\begin{equation}
{\underline r ^{\rm{f}}}\left( {\Delta k} \right) = {r^{\rm{f}}}\left( {\Delta k} \right){\mathbb{E}}\left[ {{e^{ - {{j2\pi \Delta k\theta } \over K}}}} \right],
\end{equation}
where 
$$\mathbb{E} \left[ {{e^{ - {{j2\pi \Delta k\theta } \over K}}}} \right] = \sum\limits_\theta  {p\left( \theta  \right){e^{ - {{j2\pi \Delta k\theta } \over N}}}}.$$ 

Denote $\underline {\bf{r}}^{\text{f}}$ as the vector containing ${\underline r ^{\rm{f}}}\left( {\Delta k} \right) $. With $\underline {\bf{r}}^{\text{f}}$, LMMSE estimation can achieve robust performance under different STO. $\Omega$ containing ${\mathbf{\tilde r}}^{\text{f}}_{\theta}$ for typical STO and $\underline {\bf{r}}^{\text{f}}$ is a good parameter set given consideration to performance and complexity.

In some application scenarios, the environment where systems will be used is unpredictable. Prior information about channel statistics is hard to provide for such systems. In that case, to obtain the approximation of real correlation, timely estimation methods should be adopted.  
However, the correlation estimation accuracy is influenced by many factors, and the correlation estimation result may be unacceptable when the variance of noise is extremely high and/or the pilot tones used to estimate the correlation are few. If the estimated correlation differs significantly from the true value, the performance of LMMSE estimation will degrade hugely. The estimated correlation can be replaced by robust correlation when the channel condition is bad. Therefore, the $\Omega$ constituted of estimated correlation vectors and robust correlation is a good parameter set when no prior information about CFR correlation is known.

In fact, the robust correlation is based on a coarse assumption about the maximum delay, which is normally set as the CP length. If the maximum delay is known to be shorter than half of CP length, the robust correlation can be designed as the FFT of the PDP which is uniform within $\left[ {0,...,K_{\rm{CP}}/2} \right]$, where $K_{\rm{CP}}$ is the length of CP. The performance of LMMSE estimation will be improved with the new correlation. Based on this property, $\Omega$ can be simply constituted of the robust correlation designed under different maximum delay assumptions. Here, STO can be considered in the design of $\Omega$ as well, and ${\mathbf{\tilde r}}^{\text{f}}_{\theta}$ for typical STO can be added into $\Omega$. 

\section{ Consideration on complexity}
\label{Sect.Complexity}

The computational complexity of the proposed estimator mainly depends on the number of correlation candidates $N$, and computing resource consumption is focused on the calculation of evaluation indexes for the correlation candidates. The computational complexity of the index calculation module is ${\mathcal{O}}\left({NK(K-1)^3}\right)$. It is comparable to the complexity of performing $KN$ times $(K-1)$-order LMMSE estimation. 

To reduce the complexity of proposed estimator, cutting down the number of correlation candidates is the most effective approach. But it also leads to the decrease of overall channel estimation performance. Therefore, determining the number of correlation candidates is a trade-off between performance and computational complexity. We would adopt a parameter set of modest size. 

Another approach to reduce the complexity is to simplify the calculation of the evaluation index. The  kernel of evaluation index is to calculate the difference between the interpolation value and LS estimates. In (\ref{equ.int1}), to calculate the interpolation, all the pilot tones except the interpolated one are used. Besides, the pilot tones for different interpolation always change. As a result, there are $K$ times $K-1$-order square matrix inversion in the computation of a evaluation index. In fact, to obtain the interpolation ${{\hat h}_{k\_{\rm{Int}}}}$ in (\ref{equ.metric}), (\ref{equ.int1}) is not the only approach. For the sake of reducing complexity, the number of pilot tones used in interpolation can be reduced, and the same pilot tones can be used to calculate the interpolation of different subcarriers' channel fading. To this end, we design an alternative algorithm to calculate the evaluation index. 

\begin{figure}
\begin{centering}
\includegraphics[scale=0.8]{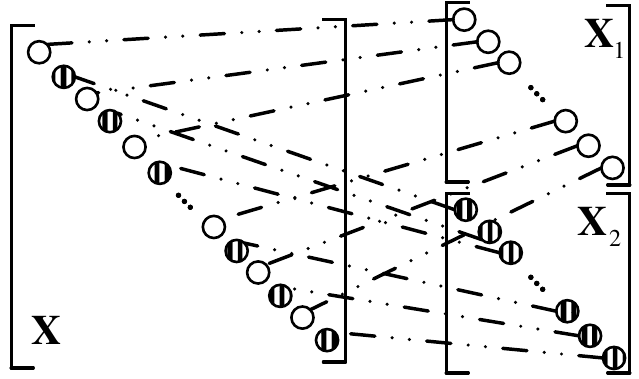}

\end{centering}
\caption{Sketch diagram for the pilot symbol division.}
\label{fig1}
\end{figure}

First, we separate the subcarriers in one pilot symbol into two groups. To guarantee good performance of interpolation,  the subcarriers between the two group should be adjacent. So we adopt an scheme that the subcarriers of the two group are uniformly spaced as shown in Fig. \ref{fig1}. We denote the pilot signal of the two group as ${{\bf{X}}_1}$ and ${{\bf{X}}_2}$ respectively. Similarly, we use ${{\bf{h}}_1}$ and ${{\bf{h}}_2}$ to represent the CFR of ${{\bf{X}}_1}$ and ${{\bf{X}}_2}$ respectively. ${{\bf{y}}_1}$ and ${{\bf{y}}_2}$ are the received signal of ${{\bf{X}}_1}$ and ${{\bf{X}}_2}$, i.e.
\begin{equation}
\begin{aligned}
   &{{\bf{y}}_1} = {{\bf{X}}_1}{{\bf{h}}_1} + {{\bf{z}}_1}  \cr 
   &{{\bf{y}}_2} = {{\bf{X}}_2}{{\bf{h}}_2} + {{\bf{z}}_2}  ,
\end{aligned}
\end{equation}
where ${\bf{z}}_1$ and ${\bf{z}}_2$ stand for noise vectors. 

We use pilot tones in ${{\bf{X}}_1}$ to calculate the interpolations of channel fading in ${{\bf{h}}_2}$ and pilot tones in ${{\bf{X}}_2}$ to calculate the interpolations of channel fading in ${{\bf{h}}_1}$. Then, the evaluation index will be
\begin{equation}
\label{equ.metric2}
\xi  = {1 \over K}\left( {\left\| {{{{\bf{\hat h}}}_{{\rm{1}}\underline \; {\rm{Int}}}} - {{{\bf{\hat h}}}_{{\rm{1}}\underline \; {\rm{LS}}}}} \right\|_2^2 + \left\| {{{{\bf{\hat h}}}_{2\underline \; {\rm{Int}}}} - {{{\bf{\hat h}}}_{2\underline \; {\rm{LS}}}}} \right\|_2^2} \right),
\end{equation}   
where ${{{\bf{\hat h}}}_{{\rm{1}}\underline \; {\rm{Int}}}}$ and ${{{\bf{\hat h}}}_{2\underline \; {\rm{Int}}}}$ are expressed as following: 
$$
\begin{aligned}
{{\bf{\hat h}}_{1\_ {\rm{Int}}}} &= {{\bf{R}}_{{{\bf{h}}_1}{{\bf{h}}_2}}}{\left( {{{\bf{R}}_{{{\bf{h}}_2}{{\bf{h}}_2}}} + {\sigma _{\rm n}^2}{\bf{I}}} \right)^{ - 1}}{{\bf{\hat h}}_{2\_ {\rm{LS}}}},\cr
{{\bf{\hat h}}_{2\_ {\rm{Int}}}} &= {{\bf{R}}_{{{\bf{h}}_2}{{\bf{h}}_1}}}{\left( {{{\bf{R}}_{{{\bf{h}}_1}{{\bf{h}}_1}}} + {\sigma _{\rm n}^2}{\bf{I}}} \right)^{ - 1}}{{\bf{\hat h}}_{1\_ {\rm{LS}}}},
\end{aligned}
$$
where ${{\bf{R}}_{{{\bf{h}}_1}{{\bf{h}}_2}}}$ is the correlation matrix between ${{\bf{h}}_1}$ and ${{\bf{h}}_2}$ and ${{\bf{R}}_{{{\bf{h}}_2}{{\bf{h}}_2}}}$ is the autocorrelation matrix of ${{\bf{h}}_2}$. ${{\bf{R}}_{{{\bf{h}}_2}{{\bf{h}}_1}}}$ and ${{\bf{R}}_{{{\bf{h}}_1}{{\bf{h}}_1}}}$ are the correlation matrix between ${{\bf{h}}_2}$ and ${{\bf{h}}_1}$ and the autocorrelation matrix of ${{\bf{h}}_1}$, respectively. ${{\bf{\hat h}}_{1\_ {\rm{LS}}}}$ and ${{\bf{\hat h}}_{2\_ {\rm{LS}}}}$ are the LS estimation of ${{\bf{h}}_1}$ and ${{\bf{h}}_2}$ respectively.

In this algorithm, there are only two $K/2$-order square matrix inversions for the calculation of each evaluation index. We can see that the complexity of algorithm is reduced significantly. However, the parameter comparison accuracy of our proposed estimator will be influenced. In this algorithm, the performance difference of the same two correlation vectors will get smaller due to the decrease of pilot tones and increase of pilot space. The shrink of performance difference will cause the decrease of parameter comparison accuracy.

To reduce the complexity further, the pilot symbol can be divided into three or more groups.  In this way, the order of square matrix inversion can be further reduced. But the reduction of complexity will be small, meanwhile the parameter comparison accuracy will decrease correspondingly. Our proposed division scheme is a good compromise between computational complexity and performance.

\section{Simulations Results and Analysis}
\label{sec.sim}

We conducted simulation experiments to verify our theoretical analyses and the performance of the proposed estimator. We built an OFDM simulation system, in which the channel coefficients are based on the models from ITU Recommendation \cite{ITUR}. Slow fading channel models, such as Office B, are mainly used, since we employ a block pilot arrangement for the OFDM system. The system parameters are shown in Table \ref{tab.para} and the channel parameters are shown in Table \ref{tab.Chpara}. The research work in this paper is focused on channel estimation, and thus only the pilot symbols are transmitted in the simulations. The time interval between the pilot symbols is normally long in practical systems, so we assume that the CFR of different pilot symbols is independent. Moreover, we consider the extreme scenario that only one pilot symbol is available in one transmission and the channel model changes randomly at the next transmission. But during one transmission, CIR is constant. In practical systems, the direct current (DC) carrier and carriers at the edges of the spectrum are normally set null. Therefore, we introduce virtual carriers in simulations, and the number of available carriers $K$ is 408 without statement.
\begin{table}  
\caption{System parameters}
\label{tab.para}  
\begin{tabular}{c|c||c|c}  
\hline  
Number of subcarriers & 512  & CP length  & 128 \\  
\hline   
\multirow{3}*{Channel models \cite{ITUR}} & Office B & Sample time  & $0.1\mu s$ \\  
\cline{3-4}
 & Pedestrian A & Symbol period  & $64\mu s$ \\  
\cline{3-4}
 & Pedestrian B & Carrier frequency  & $900\rm{MHz}$ \\
\hline
\end{tabular}  
\end{table}  

\begin{table}  
\caption{Channel parameters}
\label{tab.Chpara}  
\newcommand{\tabincell}[2]{\begin{tabular}{@{}#1@{}}#2\end{tabular}}
\begin{tabular}{c||c|c|c|c|c|c|c}  
\hline  
\multicolumn{2} {c|}{Tap} & 1  & 2  & 3 & 4 & 5 & 6 \\  
\hline   
\multirow{2}*{Office B} &  \tabincell{c}{Delay\\ (ns)} & 0  & 100 & 200 & 300 & 500 & 700 \\  
\cline{2-8}
 & \tabincell{c}{Power\\ (dB)} & 0  & -3.6 & -7.2 & -10.8 & -18.0 & -25.5 \\  
\hline
\multirow{2}*{Pedestrian A} &  \tabincell{c}{Delay\\ (ns)} & 0  & 110 & 190 & 410 & - & - \\  
\cline{2-8}
 & \tabincell{c}{Power\\ (dB)} & 0  & -9.7 & -19.2 & -22.8 & - & - \\  
\hline
\multirow{2}*{Pedestrian B} &  \tabincell{c}{Delay\\ (ns)} & 0  & 200 & 800 & 1200 & 2300 & 3700 \\  
\cline{2-8}
 & \tabincell{c}{Power\\ (dB)} & 0  & -0.9 & -4.9 & -8.0 & -7.8 & -23.9 \\  
\hline
\end{tabular}  
\end{table}  
 
\begin{figure}[!htb]
\begin{centering}
\includegraphics[scale=0.95]{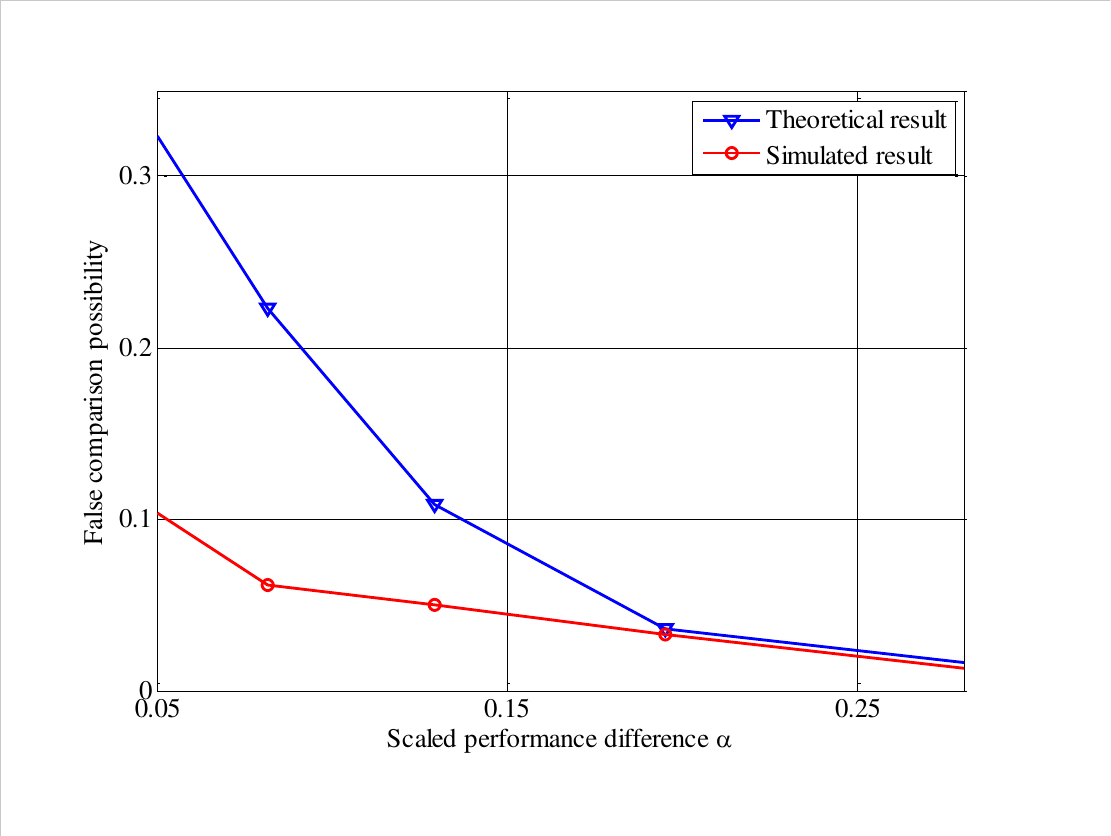}
\par\end{centering}
\caption{Error possibility varied with usable subcarriers under different SNR.}
\label{fig.TheoryComp}
\end{figure}

To verify the theoretical expression of false comparison possibility $\varepsilon $ formulated in (\ref{equ.EasyPossib}), we conduct an experiment to compare the theoretical result with the numerical result obtained from the simulated system. In the simulation experiment, $\Omega$ contains the correlation vectors for Office B and Pedestrian B, and the channel model in the simulated system is Office B. Therefore, if the correlation vector for Pedestrian B is chosen by the parameter comparison algorithm, false comparison occurs and the frequency of false comparison is the simulated result $\varepsilon $. On the other hand, by substituting the value of $\alpha$ and $K$ into (\ref{equ.EasyPossib}), the theoretical value of $\varepsilon $ can be obtained. We simulate $\varepsilon $ in the OFDM system under a SNR range from 0 dB to 8dB, since different number of $\alpha$ can be simulated by adjusting the value of SNR. The number of available carriers per symbol $K$ is set as 160. The simulation result is displayed in Fig. \ref{fig.TheoryComp}. We can see that the theoretical result is close to the simulated result when the value of $\alpha$ is high, but these two results have relatively big difference when the value of $\alpha$ is small. This difference mainly results from the two assumptions in the deduction of $\varepsilon $. This simulation result verifies our theoretical analysis about the accuracy of parameter comparison scheme to a certain degree. Although the theoretical result is not that accurate, the prediction of $\varepsilon $ using (\ref{equ.EasyPossib}) is effective, since the theoretical result can be used as the upper bound of false comparison possibility.

\begin{figure}[!htb]
\begin{centering}
\includegraphics[scale=0.9]{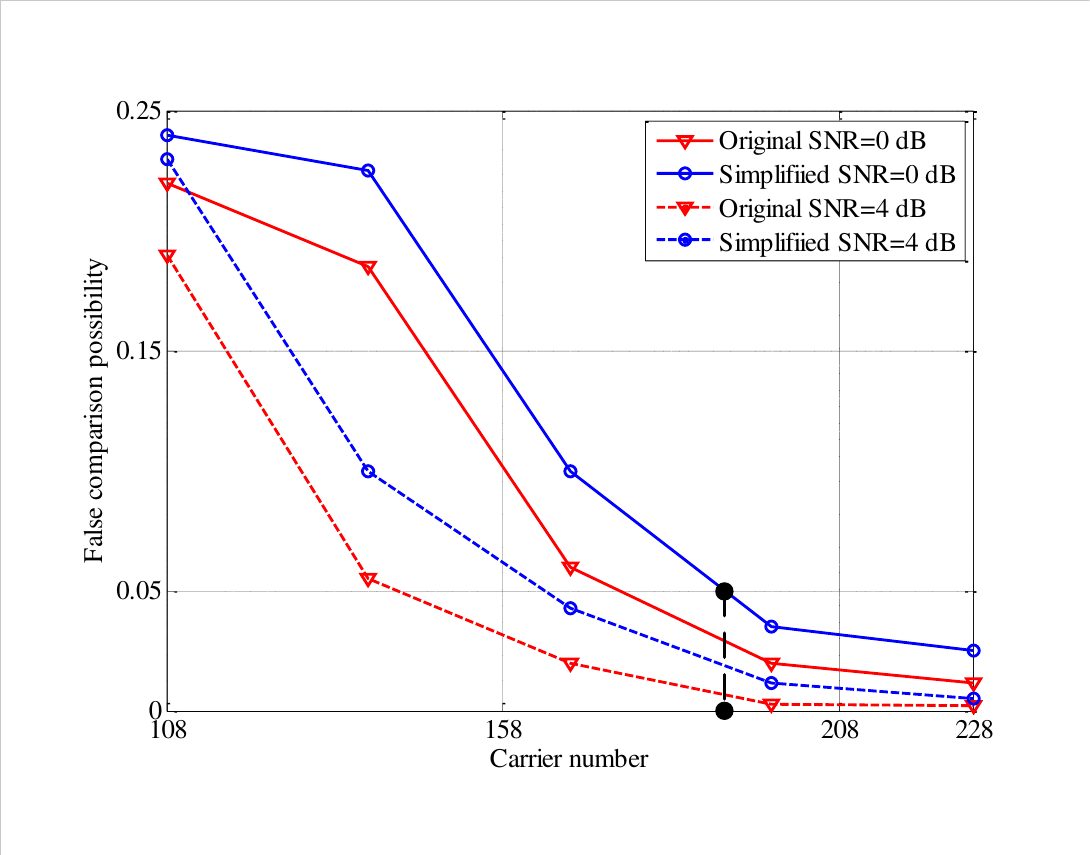}
\par\end{centering}
\caption{Error possibility varied with usable subcarriers under different SNR.}
\label{fig.possib}
\end{figure}

To examine the performance loss of the proposed simplified algorithm, we compared the accuracy of parameter selection using (\ref{equ.metric}) with that using (\ref{equ.metric2}). We simulate the error possibility varied with usable subcarriers $K$ by setting different number of null carriers at the edges of the spectrum. From Fig. \ref{fig.possib}, we can see that the parameter comparison accuracy of the simplified algorithm is a little worse than original one.   But when parameter comparison accuracy decrease is within an small range, the performance loss of the enhanced LMMSE estimator is negligible. Therefore, the complexity can be reduced without much performance loss using (\ref{equ.metric2}) to calculate the evaluation indexes. In the following simulations, we replace (\ref{equ.metric}) with (\ref{equ.metric2}) in the parameter comparison algorithm. From Fig. \ref{fig.possib}, we can also see that when usable subcarriers exceeds 200, the error possibility is below 0.05 even at low SNR. It shows that the proposed parameter comparison scheme can select the right correlation vector at high possibility, since the number of usable subcarriers is normally above 200 in practical systems.

Then, we investigated the overall estimation performance of the proposed estimator. In Section \ref{Sect.ParaSetDes}, we have described the design of parameter set for different kinds of application scenarios. Therefore, we conducted a simulation experiment for each application scenario. There are four scenarios considered: prior information about channel statistics is known; STO is present and prior information about channel and STO statistics is known; no prior information is known, but stationary period of channel is long; no prior information is known, and stationary period of channel is extremely short, which equals symbol period. 

\begin{figure}
\begin{centering}
\includegraphics[scale=0.9]{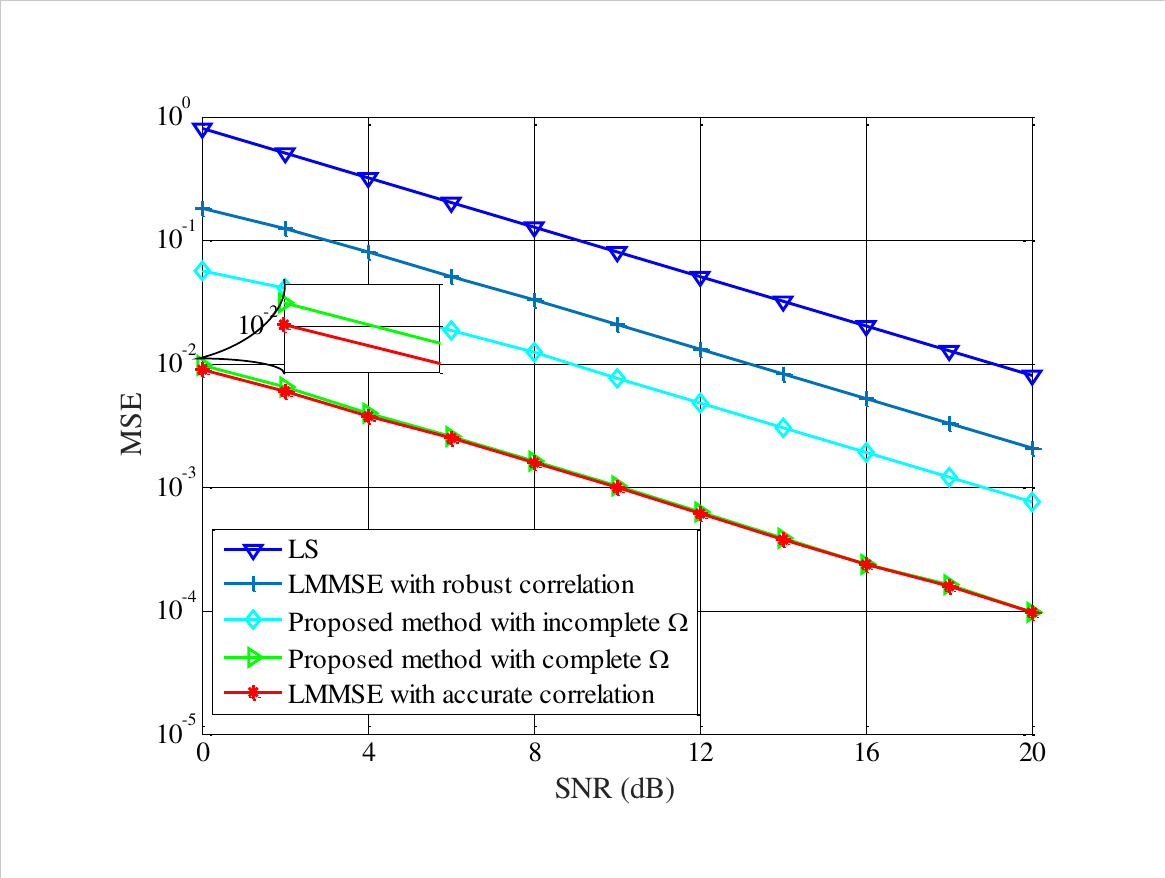}
\end{centering}
\caption{The MSE performance of proposed estimation method with complete and incomplete $\Omega$, LS estimation, and LMMSE estimation with accurate correlation information under different SNR.}
\label{fig2}
\end{figure}

In the simulation experiment for the first application scenario, we employ two kind of parameter sets in the proposed estimator: complete $\Omega$ and incomplete $\Omega$. For complete $\Omega$, we assume that the possible CFR correlation is all known. As explained in Section \ref{Sect.ParaSetDes}, $\Omega$ should contain typical correlation vectors and one robust correlation vector. Therefore, the complete $\Omega$ contains three correlation vectors which are based on the PDP of the channel models and one robust correlation vector which is based on the PDP of (\ref{equ.PDP}). The incomplete $\Omega$ contains correlation vectors for Pedestrian A and Office B and the robust correlation vector. The incomplete $\Omega$ stands for the situation that not all of the possible CFR correlation is known. From Fig. \ref{fig2}, we can see that the performance of proposed LMMSE estimation is close to that of accurate LMMSE estimation with all possible CFR correlation known. However, when $\Omega$ is not complete (missing the correlation vector for Pedestrian B), the performance of proposed method degrades but is still better than the robust LMMSE estimation. The simulation result exhibits that the performance of LMMSE estimation can be improved greatly through selecting the suitable correlation vector, and the promotion will be greater when more information about the CFR correlation is known. This simulation experiment is an ideal case for the practical systems. In practical systems, what $\Omega$ contains are the approximate correlation vectors for the real channel. Although the practical performance of proposed estimator may be worse than the simulation result, the significant performance promotion brought by parameter comparison scheme still exists in the practical systems. This property can be verified by the simulation of incomplete $\Omega$. Because incomplete $\Omega$ does not contain all the accurate correlation, and this condition is similar with the practical cases.

\begin{figure}
\begin{centering}
\includegraphics[scale=0.85]{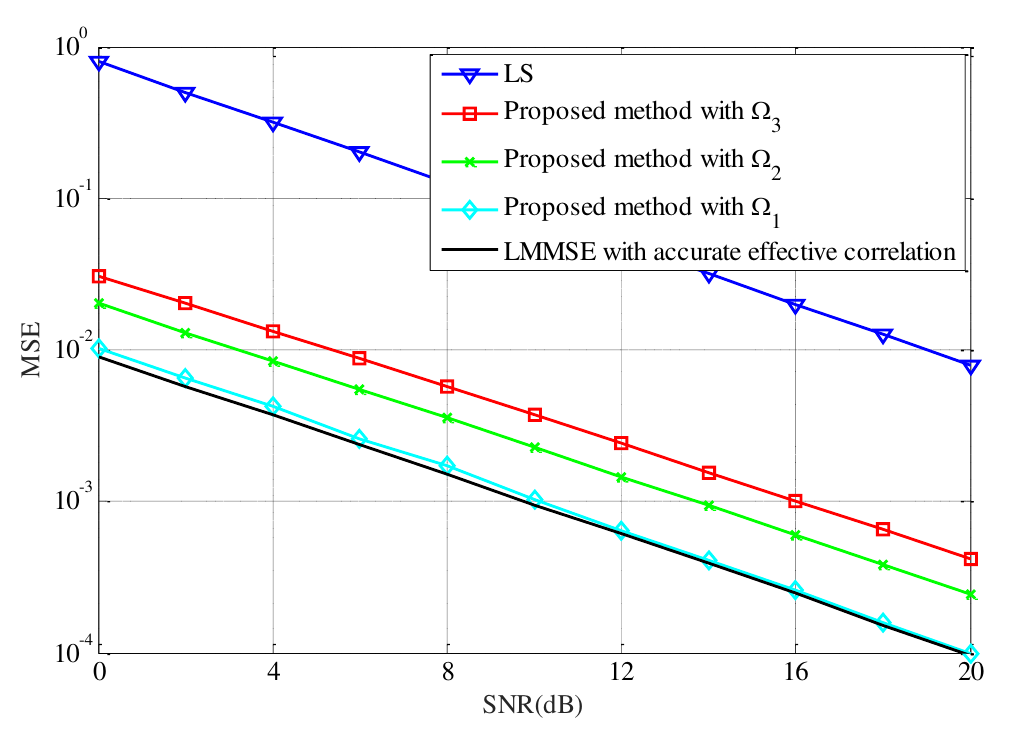}

\end{centering}
\caption{The MSE performance of proposed estimation method with three kinds of $\Omega$, LS estimation, LMMSE estimation with accurate effective correlation under different SNR with STO present.}
\label{fig3}
\end{figure}

Then, we investigate the performance of proposed method for the second scenario. We assume that the STO has equal probability within $\left[ {-10,-9,...,0} \right]$. The STO also exists in the following simulation experiments. We simulate the proposed method with three kinds of $\Omega$: $\Omega_1$ contains ${\bf{r}}_i^{\rm{f}}$ for the three models, ${\mathbf{\bar r}}_{i,\theta}^{\text{f}}$ for all STO values and $\underline {\bf{r}}_i^{\text{f}}$ calculated based on the assumed possibility distribution of STO; $\Omega_2$ contains ${\bf{r}}_i^{\rm{f}}$, ${\mathbf{\bar r}}_{i,\theta}^{\text{f}}$ for STO values from $-5$ to $-1$ and $\underline {\bf{r}}_i^{\text{f}}$; $\Omega_3$ only contains $\underline {\bf{r}}_i^{\text{f}}$. From Fig. \ref{fig3}, we can see that with more ${\mathbf{\bar r}}_{\theta}^{\text{f}}$ for specific STO values, the proposed method achieves better performance and approaches that of LMMSE estimation using accurate effective correlation when $\Omega$ contains ${\mathbf{\bar r}}_{\theta}^{\text{f}}$ for all STO values. This simulation result shows that the proposed estimator can also cope with the practical factors like STO, if these factors are considered in the design of the parameter set.

\begin{figure}
\begin{centering}
\includegraphics[scale=0.95]{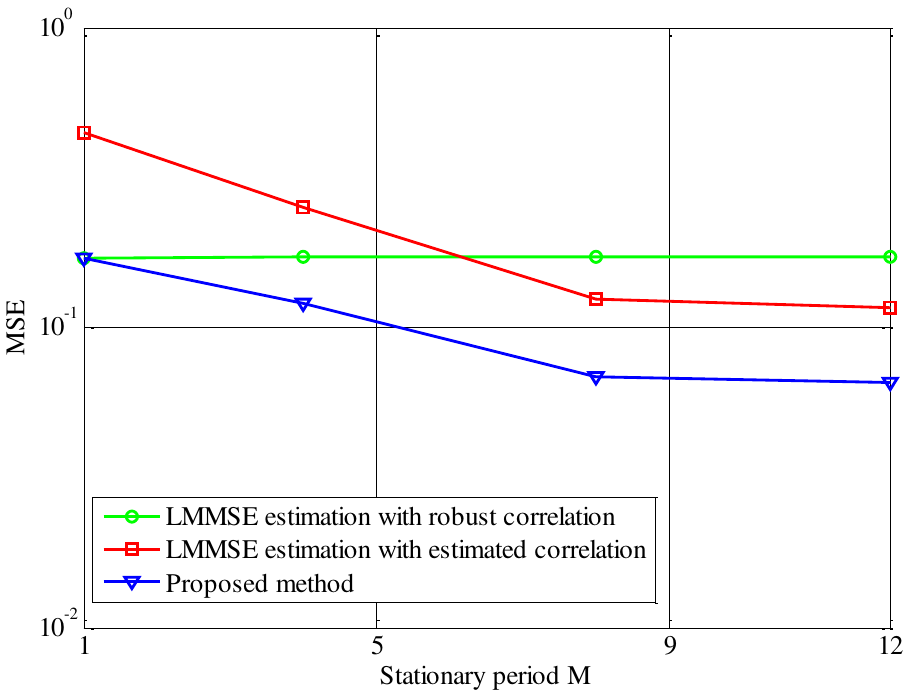}
\end{centering}
\caption{The average MSE performance of LMMSE estimation with estimated correlation, robust LMMSE and the proposed estimator under different SNR.}
\label{fig.NoPriorInf1}
\end{figure}

\begin{figure}
\begin{centering}
\includegraphics[scale=0.95]{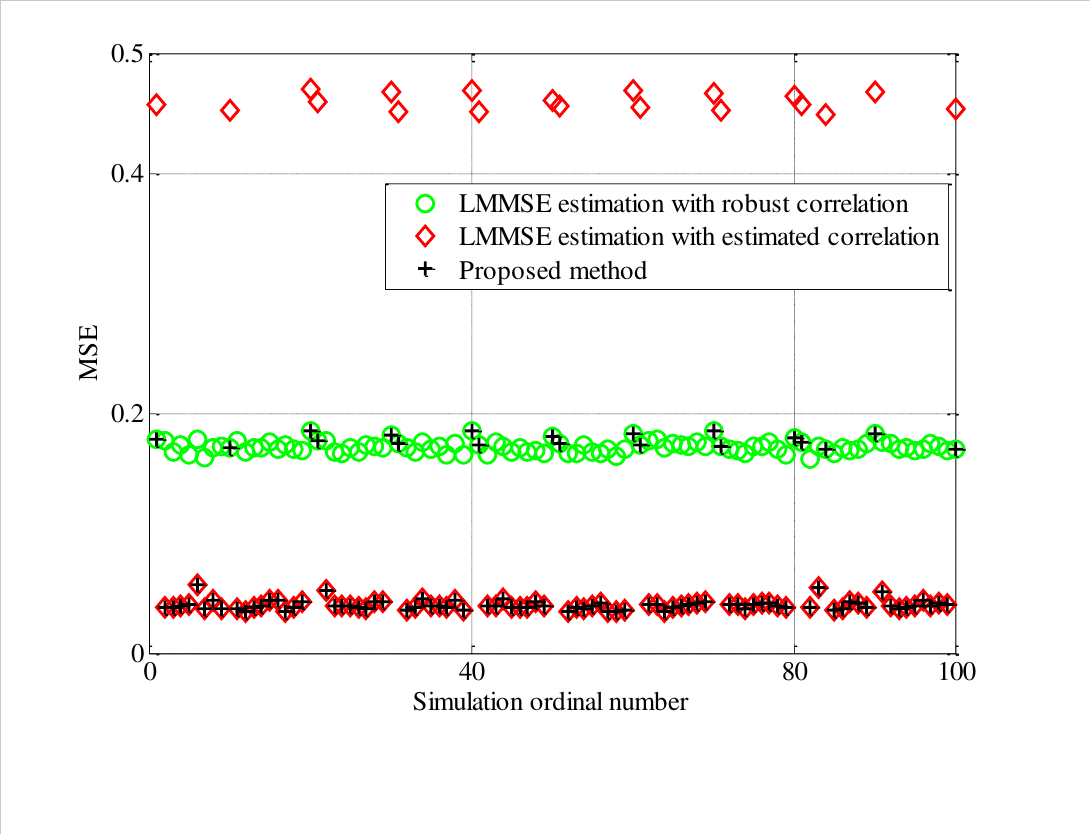}
\end{centering}
\caption{The record of MSE performance at each transmission for LMMSE estimation with estimated correlation, robust LMMSE and the proposed estimator under $M=8$.}
\label{fig.example}
\end{figure}

To demonstrate the performance of propoposed estimator under the third scenario, we simulated the system under different stationary periods. In the simulations above, the channel model changes randomly at each transmission. But in this simulation, we assume that the CFR correlation stays the same within one stationary period, and changes when the transmission time exceeds the stationary period. We use the number $M = {{T_{\rm{C}}}/{T_{\rm{S}}}}$ to indicate the stationary period, where ${T_{\rm{C}}}$ is the stationary period and ${T_{\rm{S}}}$ is the symbol period. The stationary period is assumed to be known. The parameter set is constituted of the estimated correlation vector and the robust correlation vector. We choose the approach that is based on approximate PDP to estimate the CFR correlation. The approximate PDP is a uniform model \cite{5204227}. We employ the threshold based method proposed in \cite{4167659} to estimate the scattered pathes, and then determine the delay of fisrt path $\tau_0$ and the maximum delay $\tau_{\rm{max}}$. The threshold can be expressed as
\begin{equation}
\label{equ.thre}
\lambda = {\sqrt{2} \over {M K}} \sum\limits_m{\sum\limits_k{{\hat h_{k,m}^{\rm{t}}}\left({\hat h_{k,m}^{\rm{t}}}\right)^*}},
\end{equation}
where ${\hat h_{k,m}^{\rm{t}}}$ represents the estimates of channel impulse reponse (CIR) at the $k\rm{th}$ sampled time in the $m\rm{th}$ OFDM symbol. SNR is set to 0 dB. In Fig. \ref{fig.NoPriorInf1}, the average MSE performance of LMMSE estimation with estimated correlation, robust correlation and selected correlation respectively is displayed. We can see that when stationary period is extremely short, the average MSE of proposed estimator is close to that of robust LMMSE. In fact, the MSE performance of LMMSE estimation with estimated correlation is not stable. In one transmission, its performance may be better or worse than that of the robust LMMSE. When the pilot symbols used for the estimation of correlation is few, the performance of LMMSE estimation with the estimated correlation is always worse than the robust LMMSE estimation, and the estimated correlation would not be selected in the proposed estimator. Therefore, the average MSE of proposed estimator and robust LMMSE will be the same. With the growing of stationary period, the performance of LMMSE estimation with estimated CFR correlation will be better than robust LMMSE more frequently. Thus, the average MSE of LMMSE estimation with estimated CFR gradually exceeds that of robust LMMSE with the increase of stationary period. Meanwhile, the average MSE of the proposed estimator decreases correspondingly. This phenomenon is illustrated in Fig. \ref{fig.example}. It is the record of MSE at every transmission, when stationary period is $8$. We can see that the MSE of LMMSE estimation with estimated correlation jumps between two main intervals, and the proposed estimator always achieves the better performance between the two conventional LMMSE estimation methods. Therefore, our proposed estimator has the lowest average MSE.

\begin{figure}
\begin{centering}
\includegraphics[scale=0.95]{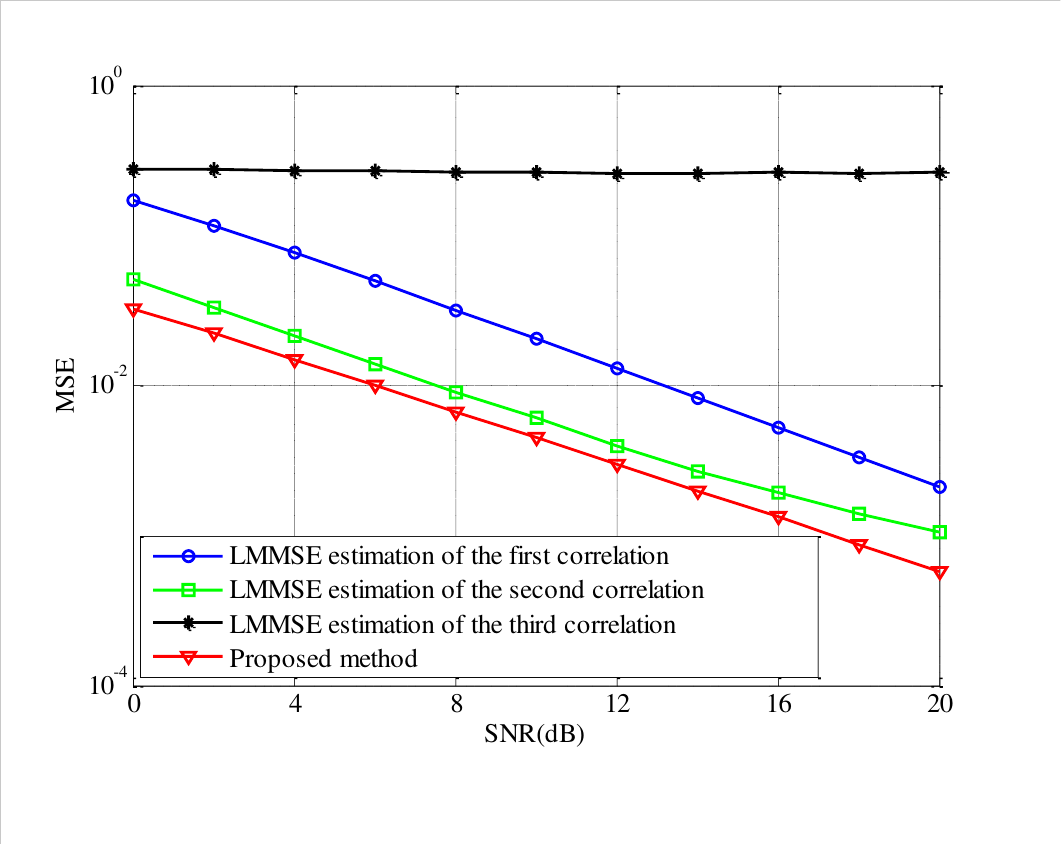}
\end{centering}
\caption{The MSE performance of proposed estimation method, and LMMSE estimation with $\tau _{\rm{max}}$ set to $K_{\rm{CP}}$, $K_{\rm{CP}}/4$ and $K_{\rm{CP}}/16$ under different SNR with STO present.}
\label{fig.NoPriorInf2}
\end{figure}

Under the fourth scenario, it is difficult to obtain usable CFR correlation timely. The parameters of LMMSE estimation should be determined in advance. Therefore, the parameter set is constituted of correlation vectors based on the PDP in (\ref{equ.PDP}) with $\tau _{\rm{max}}$ set to $K_{\rm{CP}}$, $K_{\rm{CP}}/4$ and $K_{\rm{CP}}/16$ respectively and ${\mathbf{\bar r}}_{i,\theta}^{\text{f}}$ for STO values from $-5$ to $-1$. We simulated the proposed estimation under different SNR. In Fig. \ref{fig.NoPriorInf2}, the first correlation, second correlation and third correlation are FFT of PDP with $\tau _{\rm{max}}$ set to $K_{\rm{CP}}$, $K_{\rm{CP}}/4$ and $K_{\rm{CP}}/16$ respectively. We can see that although there is no prior information about channel and the channel condition is harsh, the proposed estimator can still improve the estimation performance significantly. Without the parameter comparison scheme, only the robust correlation can be used under such channel condition, although its performance is not good. The LMMSE estimation with $\tau _{\rm{max}}$ set to $K_{\rm{CP}}/16$ may have much better performance if the real maximum delay is below $K_{\rm{CP}}/16$, but it has the worst performance, since the maximum delay is above $K_{\rm{CP}}/16$ in the simulation. This phenomenon exhibits the common problem for some of conventional methods: when the channel assumptions mismatch the real channel, these methods may suffer huge performance degradation. With the parameter comparison scheme, this problem can be well solved.

\section{Conclusions}
 \label{sec.5}
In this paper, we present an enhanced LMMSE estimator based on a novel parameter comparison scheme. The analysis results show that the parameter comparison scheme can always promote the average performance of LMMSE estimation. The false comparison possibility is deduced in a closed form, and thus the parameter comparison accuracy is predictable when the length of sequence is given. The proposed LMMSE estimator is then applied in OFDM systems with the block pilot arrangement. To this end, we designed a LMMSE estimator capable of selecting CFR correlation. We also illustrated the design of the parameter set under two types of conditions: prior information about channel is known; there is no prior information available. Besides, we investigated the computational complexity of proposed estimator and propose an estimator with much lower complexity. Simulation results verify our theoretical analysis and the performance of the proposed  estimator. The simplified algorithm has been compared with the original proposed estimator, and the simulation result shows that the performance loss caused by the simplification of calculation is slight. The comparison between our proposed estimator and the traditional ones exhibits the significant performance promotion brought by the proposed parameter comparison scheme. The simulation experiments in which STO is considered exhibits the proposed estimator's adaptability for practical systems.
 
% or

\appendices
\section{Deduction of $\mathbb{E}\left[  {{\left| {{{\hat h}_{k\_{\rm{Int}}}} - {{\hat h}_{k\_{\rm{LS}}}}} \right|}^2} \right]$}

\label{IndexVariance}

$$
\begin{gathered}
  \mathbb{E}\left[ {{{\left| {{{\hat h}_{k\_{\rm{Int}}}} - {{\hat h}_{k\_{\rm{LS}}}}} \right|}^2}} \right] \hfill \\
   = \mathbb{E}\left[ {\left( { {{\hat h}_{k\_{\rm{Int}}}} - {{\hat h}_{k\_{\rm{LS}}}}} \right){{\left( { {{\hat h}_{k\_{\rm{Int}}}} - {{\hat h}_{k\_{\rm{LS}}}}} \right)}^ * }} \right] \hfill \\
   = \mathbb{E}\left[ {\left( { {{\hat h}_{k\_{\rm{Int}}}} - {h_k} -{n_k} } \right)}  {{{\left( { {{\hat h}_{k\_{\rm{Int}}}} - {h_k} -{n_k} } \right)}^ * }} \right] \hfill \\ 
   = \mathbb{E}\left[ {\left( { {{\hat h}_{k\_{\rm{Int}}}} - {h_k} } \right)}  {{{\left( { {{\hat h}_{k\_{\rm{Int}}}} - {h_k}  } \right)}^ * }} \right] \hfill \\ 
   \;\;\;\; -\mathbb{E}\left[ {\left( { {{\hat h}_{k\_{\rm{Int}}}} - {h_k} } \right)}  {{{\left( { n_k } \right)}^ * }} \right] \hfill \\
     \;\;\;\; -\mathbb{E}\left[ {\left( { {{\hat h}_{k\_{\rm{Int}}}} - {h_k} } \right)}^ *  { n_k } \right] +\mathbb{E}\left[   { n_k } {\left( { n_k } \right)}^ * \right], \hfill \\
\end{gathered}  
$$
where $n_k$ is the $k\rm{th}$ element of white noise vector $\bf{n}$ and independent of any other random variables. ${{\hat h}_{k\_{\rm{LS}}}}$ is not used in ${{\hat h}_{k\_{\rm{Int}}}}$, so $n_k$ is independent of ${{\hat h}_{k\_{\rm{Int}}}}$. Thus, $\mathbb{E}\left[ {\left( { {{\hat h}_{k\_{\rm{Int}}}} - {h_k} } \right)}^ *  { n_k } \right] = 0$ and $\mathbb{E}\left[ {\left( { {{\hat h}_{k\_{\rm{Int}}}} - {h_k} } \right)}  {{{\left( { n_k } \right)}^ * }} \right] = 0$. The above formula can be simplified as
$$
\begin{gathered}
  \mathbb{E}\left[ {{{\left| {{{\hat h}_{k\_{\rm{Int}}}} - {{\hat h}_{k\_{\rm{LS}}}}} \right|}^2}} \right] \hfill \\
   = \mathbb{E}\left[ {\left( { {{\hat h}_{k\_{\rm{Int}}}} - {h_k} } \right)}  {{{\left( { {{\hat h}_{k\_{\rm{Int}}}} - {h_k}  } \right)}^ * }} \right] \hfill \\
  \;\;\;\; + \mathbb{E}\left[   { n_k } {\left( { n_k } \right)}^ * \right] \hfill \\
   = \sigma _{{\text{MSE}}}^2 + \sigma _{\rm{LS}}^2 , \hfill \\ 
\end{gathered}  
$$
where $\sigma _{{\text{MSE}}}^2$ represents the MSE of interpolation.

\section{Possibility of false comparison}
\label{Possibility}

$$
\begin{gathered}
  \mathbb{E} \left[ {{{\hat h}_{k\_{\rm{Int}}}} - {{\hat h}_{k\_{\rm{LS}}}}} \right] = \mathbb{E} \left[ {{{\hat h}_{k\_{\rm{Int}}}}} \right] - \mathbb{E} \left[ {{{\hat h}_{k\_{\rm{LS}}}}} \right] \hfill \\
   = \mathbb{E} \left[ {{{\bf{r}}_{{h_k}{{\bf{h}}_{k{\rm{Ex}}}}}}{{\left( {{{\bf{R}}_{{{\bf{h}}_{k{\rm{Ex}}}}{{\bf{h}}_{k{\rm{Ex}}}}}} + \sigma _{{\rm{LS}}}^2{\bf{I}}} \right)}^{ - 1}}{{{\bf{\hat h}}}_{k{\rm{Ex}}\_{\rm{LS}}}}} \right]  \hfill \\
  \;\;\;\; -  \left( {\mathbb{E} \left[ {{h_k}} \right] + \mathbb{E} \left[ {{n_k}} \right]} \right) \hfill \\
   = {{\bf{r}}_{{h_k}{{\bf{h}}_{k{\rm{Ex}}}}}}{\left( {{{\bf{R}}_{{{\bf{h}}_{k{\rm{Ex}}}}{{\bf{h}}_{k{\rm{Ex}}}}}} + \sigma _{{\rm{LS}}}^2{\bf{I}}} \right)^{ - 1}}\left( {\mathbb{E} \left[ {{{\bf{h}}_{k{\rm{Ex}}}}} \right] + \mathbb{E} \left[ {{{\bf{n}}_{k{\rm{Ex}}}}} \right]} \right) \hfill \\ 
   = 0. \hfill \\ 
\end{gathered}  
$$

The expectation of $\left( {{{\hat h}_{k\_{\rm{Int}}}} - {{\hat h}_{k\_{\rm{LS}}}}} \right) $ is 0, so the variance of $\left( {{{\hat h}_{k\_{\rm{Int}}}} - {{\hat h}_{k\_{\rm{LS}}}}} \right)$ equals its autocorrelation derived in Appendix \ref{IndexVariance}, i.e. $\mathbb{D} \left[ {{{\hat h}_{k\_{\rm{Int}}}} - {{\hat h}_{k\_{\rm{LS}}}}} \right] = \sigma _{{\text{MSE}}}^2 + \sigma _{\rm{LS}}^2$. Since the linear combination of complex Gaussian random is also complex Gaussian random and the random variables in $\left( {{{\hat h}_{k\_{\rm{Int}}}} - {{\hat h}_{k\_{\rm{LS}}}}} \right) $ are all complex Gaussian variables, $\left( {{{\hat h}_{k\_{\rm{Int}}}} - {{\hat h}_{k\_{\rm{LS}}}}} \right) $ is subject to complex Gaussian distribution, i.e. $\left( {{{\hat h}_{k\_{\rm{Int}}}} - {{\hat h}_{k\_{\rm{LS}}}}} \right) \sim {\cal{CN}}\left( {0,\sigma _{{\rm{MSE}}}^2 + \sigma _{{\rm{LS}}}^2} \right)$.

Denote the evaluation index of ${\pmb{\theta} _1}$ and ${\pmb{\theta} _2}$ as $\xi_1$ and $\xi_2$ respectively. With the assumption that the interpolations' noise MSE  using ${\pmb{\theta} _1}$ and ${\pmb{\theta} _2}$ are $\sigma _1^2$ and $\sigma _2^2$ respectively, it can be easily derived that $\left( {{{\hat h}_{k\_{\rm{Int}}\left| {{{\pmb{\theta }}_1}} \right.}} - {{\hat h}_{k\_{\rm{LS}}}}} \right)\sim {\cal{CN}} \left( {0,\sigma _{\rm{1}}^2 } \right)$ and $\left( {{{\hat h}_{k\_{\rm{Int}}\left| {{{\pmb{\theta }}_2}} \right.}} - {{\hat h}_{k\_{\rm{LS}}}}} \right)\sim {\cal{CN}} \left( {0,\sigma _2^2 } \right)$.

For simplification, we make two assumptions: first, $\left( {{{\hat h}_{{k_1}\_{\rm{Int}}}} - {{\hat h}_{{k_1}\_{\rm{LS}}}}} \right)$ is independent of $\left( {{{\hat h}_{{k_2}\_{\rm{Int}}}} - {{\hat h}_{{k_2}\_{\rm{LS}}}}} \right)$, when ${k_1} \ne {k_2}$; second, $\xi_1$ is independent of $\xi_2$.

Under the first assumption, the scaled indexes $2K\xi_1 / { \sigma _1^2} $ and $2K\xi_2 / { \sigma _2^2}$ are both subject to the chi-square distribution ${\chi ^2}\left( {2K} \right)$. Let ${\kappa} = 2K$ be the dimension of the chi-square distribution, and the possibility distribution function (PDF) of $\xi_1$ is 
$$
p_1 \left( x  \right) = {\kappa \over { \sigma _1^2}}{p_{\chi _{\kappa}^2}}\left( {{{\kappa x } \over { \sigma _1^2}}} \right)
$$
where ${p_{\chi _{\kappa}^2}}\left(  \cdot  \right)$ is the possibility density function of ${\chi ^2}\left( {\kappa} \right)$. The PDF of  $\xi_2$ is 
$$
p_2 \left( x  \right) = {\kappa \over { \sigma _2^2}}{p_{\chi _{\kappa}^2}}\left( {{{\kappa x } \over { \sigma _2^2}}} \right)
$$

False comparison occurs when ${\xi _1} \ge {\xi _2}$, so the false comparison possibility is $\varepsilon  = P\left( {{\xi _1} \ge {\xi _2}} \right)$. Assume that ${p\left( {{x_1},{x_2}} \right)}$ is the joint PDF of $\xi_1$ and $\xi_2$, $\varepsilon$ can be expressed as $\varepsilon = \int_0^\infty  {\int_0^{{x_1}} {p\left( {{x_1},{x_2}} \right)} } d{x_1}d{x_2}$. Under the second assumption, $p\left( {{x_1},{x_2}} \right) = p_1\left( {{x_1}} \right) \cdot p_2\left( {{x_2}} \right)$. Therefore, the expression of $\varepsilon$ can be transformed as
$$
\begin{aligned}
  \varepsilon  & = \int_0^\infty  {{p_1}\left( {{x_1}} \right)} \int_0^{{x_1}} {{p_2}\left( {{x_2}} \right)d{x_1}d{x_2}}   \cr 
  &  = \int_0^\infty  {{\kappa  \over {\sigma _1^2}}{P_{\chi _\kappa ^2}}\left( {{{\kappa {x_1}} \over {\sigma _1^2}}} \right)} \int_0^{{x_1}} {{\kappa  \over {\sigma _2^2}}{P_{\chi _\kappa ^2}}\left( {{{\kappa {x_2}} \over {\sigma _2^2}}} \right)d{x_2}d{x_1}}   \cr 
  & \mathop  {=\hfill} \limits^{{\zeta _2} = {{\kappa {x_2}} \over {\sigma _2^2}}} \int_0^\infty  {{\kappa  \over {\sigma _1^2}}{P_{\chi _\kappa ^2}}\left( {{{\kappa {x_1}} \over {\sigma _1^2}}} \right)} \int_0^{{{\kappa {x_1}} \over {\sigma _2^2}}} {{P_{\chi _\kappa ^2}}\left( {{\zeta _2}} \right)d{\zeta _2}d{x_1}}   \cr 
  &  = \int_0^\infty  {{\kappa  \over {\sigma _1^2}}{P_{\chi _\kappa ^2}}\left( {{{\kappa {x_1}} \over {\sigma _1^2}}} \right){F_{\chi _\kappa ^2}}\left( {{{\kappa {x_1}} \over {\sigma _2^2}}} \right)d{x_1}}   \cr 
  & \mathop  {=\hfill} \limits^{{\varsigma _1}  = {{\kappa {x_1}} \over {\sigma _1^2}}} \int_0^\infty  {{P_{\chi _\kappa ^2}}\left( {{\varsigma _1}} \right)} {F_{\chi _\kappa ^2}}\left( {{{\sigma _1^2} \over {\sigma _2^2}}{\varsigma _1}} \right)d{\varsigma _1}  \cr 
  &  = \int_0^\infty  {{F_{\chi _\kappa ^2}}\left( {\left( {1 - {{{\Delta _{{\rm{MSE}}}}} \over {\sigma _2^2}}} \right){\varsigma _1}} \right){P_{\chi _\kappa ^2}}\left( {{\varsigma _1}} \right)} d{\varsigma _1} 
\end{aligned}
$$
where ${F_{\chi _{\kappa}^2}}\left(  \cdot  \right)$ is the CDF of ${\chi ^2}\left(\kappa\right)$. 


%\section*{Acknowledgment}


\ifCLASSOPTIONcaptionsoff
  \newpage
\fi



%\bibliographystyle{IEEEtran}
%\bibliographystyle{unsrt}
\bibliographystyle{IEEEtran}
\bibliography{Paper2}


%\end{IEEEbiography}
%\newpage


%\vfill


\end{document}